\documentclass[format=acmsmall, review=false, screen=true]{acmart}
\usepackage{hhline}
\usepackage{siunitx}
\usepackage{longtable}
\usepackage{array,multirow,adjustbox}
\usepackage{etoolbox}
\usepackage[utf8]{inputenc}
\usepackage[T1]{fontenc}
\usepackage[ruled,vlined,linesnumbered,noend]{algorithm2e}

\newenvironment{algo}[1][htb]
  {
  
   \begin{algorithm}[#1]
  }{\end{algorithm}}
  
 \AtBeginEnvironment{algo}{}
 \SetAlFnt{\small\ttfamily}
\SetAlCapFnt{\small\ttfamily}
\SetKwSty{small ttfamily}
\SetFuncSty{small ttfamily}
\SetFuncArgSty{small ttfamily}
\SetProgSty{small ttfamily}
\SetArgSty{small ttfamily}
\SetDataSty{small ttfamily}
\SetCommentSty{small ttfamily}
\SetProcNameSty{small ttfamily}
\SetProcArgSty{small ttfamily}
\SetBlockMarkersSty{small ttfamily}
\definecolor{darkgreen}{rgb}{0,0.4,0}
\usepackage{algorithmic}
\usepackage{setspace} 
\usepackage{listings}

\lstdefinelanguage{OCL}
{
morekeywords={operation, body, let, in},
sensitive=false,
morecomment=[l]{//},
morecomment=[s]{/*}{*/},
morestring=[b]",
}

\definecolor{light-gray}{gray}{1}
\definecolor{cverbbg}{gray}{0.93}
\usepackage[framemethod=TikZ]{mdframed}
\mdfsetup{skipabove=5pt,skipbelow=3pt}

\mdfdefinestyle{MyFrame}{
    linecolor=black,
    outerlinewidth=0.15pt,
    roundcorner=2pt,
    innertopmargin=2pt,
    innerbottommargin=2pt,
    innerrightmargin=2pt,
    innerleftmargin=2pt,
    backgroundcolor=cverbbg}

\newcommand{\sectopic}[1]{\vspace*{.5em}\par\noindent{\textit{\bfseries #1}}}
\newcommand{\ocl}[1]{{\fontsize{8.7}{8.7}\selectfont {\ttfamily\bfseries #1}}}
\newcommand{\var}[1]{{\it{\ttfamily\it\it #1}}}
\newcommand{\makeBlue}[1]{{\color{blue}#1}}
\newcommand{\makePurple}[1]{{\color{purple}#1}}
\newcommand{\makeBlack}[1]{{\color{black}#1}}
\newcommand{\variable}[1]{{$\langle$#1$\rangle$}}
\newcommand{\ourMid}[0]{\,$\mid$\,}

\usepackage[linesnumbered,ruled,vlined]{algorithm2e}
\let\oldnl\nl
\newcommand{\nonl}{\renewcommand{\nl}{\let\nl\oldnl}}

\newcommand\redsout{\bgroup\markoverwith{\textcolor{red}{\rule[0.5ex]{2pt}{0.4pt}}}\ULon}

\begin{document}
\title[Practical Constraint Solving for Generating System Test Data]
{Practical Constraint Solving for Generating \\ System Test Data}

\author{Ghanem Soltana}
\affiliation{\institution{University of Luxembourg}}

\author{Mehrdad Sabetzadeh}
\affiliation{\institution{School of EECS, University of Ottawa \& SnT, University of Luxembourg}}

\author{Lionel C. Briand}
\affiliation{ \institution{School of EECS, University of Ottawa \& SnT, University of Luxembourg}}

\begin{abstract}
The ability to generate test data is often a necessary prerequisite for automated software testing. For the generated data to be fit for its intended purpose, the data usually has to satisfy various logical constraints. When testing is performed at a system level, these constraints tend to be complex and are typically captured in expressive formalisms based on first-order logic. Motivated by improving the feasibility and scalability of data generation for system testing, we present a novel approach, whereby we employ a combination of metaheuristic search and Satisfiability Modulo Theories (SMT) for constraint solving. Our approach delegates constraint solving tasks to metaheuristic search and SMT in such a way as to take advantage of the complementary strengths of the two techniques. We ground our work on test data models specified in UML, with OCL used as the constraint language. We present tool support and an evaluation of our approach over three industrial case studies. The results indicate that, for complex system test data generation problems, our approach presents substantial benefits over the state of the art in terms of applicability and scalability.
\end{abstract}


\thanks{This work is supported by the European Research Council under the European Union's Horizon 2020 research and innovation program (grant agreement number 694277).}

\maketitle

\section{Introduction}\label{sec:introduction}
Test data generation is one of the most important and yet most challenging aspects of software test automation~\cite{Korel:90,McMinn:04,AliBHP10,Myers:2011}. To be useful, test data often needs to satisfy logical constraints. This makes constraint solving an integral part of test data generation. 
For example, in white-box testing, one may be interested in achieving path or state coverage. Doing so requires solving the path constraints for the different control paths in a program~\cite{McMinn:04}. Similarly, in black box testing,  one might require solving complex constraints to generate test data that comply with the system specifications. Otherwise,  trivially invalid data inputs would be frequently and easily discarded by the system under test.

Test data tends to get progressively more complex as one moves up the testing ladder, from unit to integration to system testing. Particularly, system testing, whose goal is ensuring that a system is in compliance with its requirements~\cite{Ammann:16}, involves exercising end-to-end system behaviors. This, in turn, typically requires numerous interdependent data structures, and thus the ability to account for the well-formedness and semantic correctness constraints of these data structures.

For example, system testing of an application that calculates citizens' taxes would require meaningful assemblies of many inter-related entities, including taxpayers, incomes, dependent family members, and so on. When available, real data, e.g., real tax records in the above example, may be utilized for system testing. In many practical settings, however, real data is (1) incomplete, meaning that the data is insufficient for exercising every system behavior that needs to be tested, 
(2)~incompatible, meaning that, due to changes in the system, the real data no longer matches the system's data structures, or (3)~inaccessible, meaning that the real data cannot be used for testing due to reasons such as data protection and privacy.

Because of these factors, system testing is done, by and large, using synthetic data. Generating synthetic data for system testing necessitates constraint solving over system concepts, including concepts related to a system's environment~\cite{Iqbal:15}. The languages used for expressing system-level constraints are feature-rich. Languages based on first-order logic are particularly common, noting that quantification, predicates and functions are usually inevitable when one attempts to specify constraints at a system level. This high degree of expressiveness often comes at the cost of making constraint solving undecidable in general, and computationally  expensive even when the size of the data to generate is bounded~\cite{Libkin:2004,JacksonBook2012}.

Our work in this article is prompted by the need to devise a \emph{practical} solution for generating system test data based on a set of complex constraints.
To this end, we propose a novel constraint solving approach. We ground our approach on UML~\cite{UML} and its constraint language, OCL~\cite{OCL}. This choice is motivated by UML's widespread use in industry. OCL supports, among other features, quantification and a variety of arithmetic, string and collection operations. 

Our approach leverages earlier strands of work where (metaheuristic) search is applied for solving OCL constraints~\cite{Shaukat:2013,Shaukat:2016,Soltana:2017}. 
In the context of testing, these earlier strands were motivated by the limitations -- in terms of scalability and the UML/OCL constructs covered -- of commonly used exhaustive approaches based on SAT solving, e.g., Alloy~\cite{JacksonBook2012}, and traditional constraint programming, e.g., UMLtoCSP~\cite{UMLtoCSP}. 

The core idea behind our approach is to further enhance the performance of search-based OCL solving by combining it with Satisfiability Modulo Theories (SMT)~\cite{SMT}. Specifically, we observe that SMT solvers, e.g., Z3~\cite{z3}, have efficient decision procedures for several background theories, e.g., {linear arithmetic}~\cite{dechter2003constraint}. While these background theories per se are not sufficient for solving complex OCL constraints, the following prospect exists: if a subformula $f$ in a constraint is solvable by a combination of background theories, then the associated decision procedures are likely to handle $f$ more efficiently than search. What we set out to do in this article is to provide, based on the above prospect, a fully fledged combination of search and SMT for OCL solving.

\vspace*{.3em}\sectopic{Contributions.} The contributions of this article are three-fold:

\vspace*{.3em}\noindent{\emph 1)}  We develop a \emph{hybrid} approach for solving OCL constraints using search and SMT. The approach first transforms OCL constraints into a normal form, known as the Negation Normal Form (NNF)~\cite{NNFBook}. NNF is key to our approach by enabling an effective distribution of constraint solving tasks across search and SMT. Although our focus in this article remains on OCL, the principles and practical considerations underlying our solving strategy are general,  and can be applied to other constraint languages. 

\vspace*{.3em}\noindent{\emph 2)} We evaluate our approach on \emph{three industrial case studies} from distinct application domains. {The case studies are all concerned with the generation of data for system testing. In each case, we automatically produce synthetic test data of varying sizes.} Our results indicate that our approach is able to solve complex OCL constraints while generating, in practical time, fairly large data samples. We further compare against alternative solvers, and demonstrate that our approach leads to significant improvements in applicability and scalability. It is important to note that our evaluation is not aimed at assessing the effectiveness of different test strategies, but rather at assessing the feasibility and efficiency of automatically generating system test data regardless of test strategy.

{\vspace*{.1em}\noindent{\emph 3)} In support of our approach, we develop a tool named  PLEDGE (PracticaL and Efficient Data GEnerator for UML). The tool is publicly available at 
\url{https://sites.google.com/view/hybridoclsolver/}}. To facilitate the replication of our empirical results and also to support future research on test data generation, we provide on the above website our (sanitized) case study material as well.

\sectopic{Structure.}
Section~\ref{sec:example} presents a running example. 
Section~\ref{sec:background} provides background.
Section~\ref{sec:approach} outlines our approach.
Sections~\ref{sec:ocl} and \ref{sec:solving} elaborate the technical aspects of the approach. 
Section~\ref{sec:evaluation} describes our evaluation. 
Section~\ref{sec:threats} discusses limitations and threats to validity. 
Section~\ref{sec:related} compares with related work.
Section~\ref{sec:conclusion} concludes the article.

\section{Running Example}\label{sec:example}
\begin{figure}[!t]
	\centering
	\includegraphics[width=0.75\linewidth]{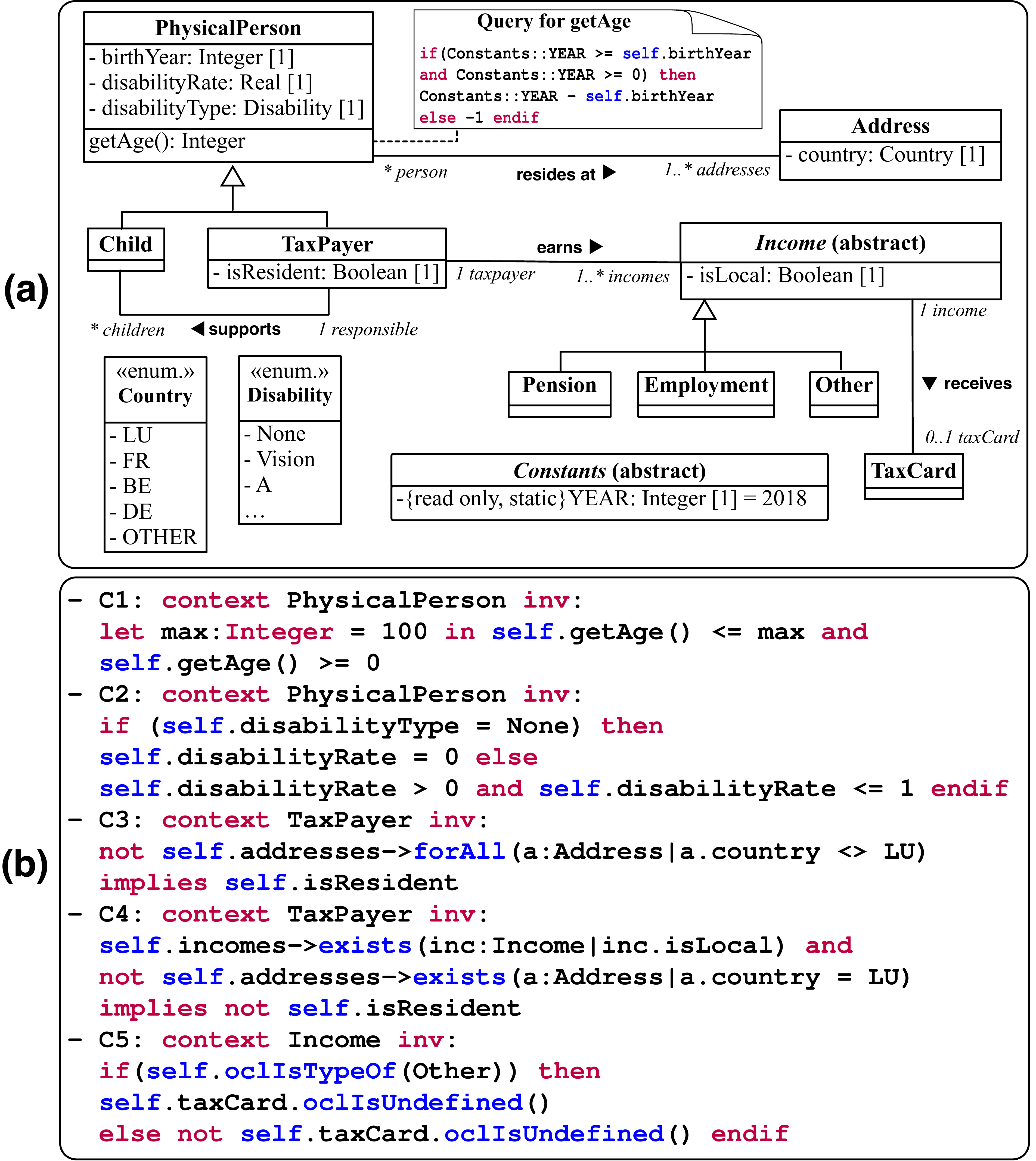}
\caption{{ (a)  Example UML Class Diagram, and (b) Example OCL Constraints \label{fig:example}}}
\end{figure}

In this section, we present an example which we use throughout the article for illustration. 
For several practical reasons, system test generation is usually black-box and requires an explicit test data model~\cite{Utting:07}. 
In UML, test data models can be  expressed with relative ease using Class Diagrams (CD) augmented with OCL constraints~\cite{Warmer:03}. The example, shown in Fig.~\ref{fig:example}, is a small and simplified excerpt of the system test data model for the Government of Luxembourg's personal income tax management application. The full data model is the subject of one of our case studies in Section~\ref{sec:evaluation}.

The graphical notation of a CD already places certain constraints on how the classes and associations in a CD can be instantiated~\cite{UML}. One basic constraint is that abstract classes, e.g., \textit{Income} in the CD of Fig.~\ref{fig:example}(a), cannot be instantiated. A second basic constraint is that the objects on the two ends of a link (association instance) must conform to the CD. For example, a \textit{supports} link should have a \textit{Taxpayer} object on one end and a \textit{Child} object on the other.  A CD may additionally have multiplicity constraints. For example, the multiplicities attached to the two ends of the \textit{earns} association impose the following constraints: (1) each \textit{TaxPayer} object must be linked to at least one object of kind \textit{Income}; and (2) each object of kind \textit{Income} must be linked to \hbox{exactly one \textit{TaxPayer} object.}

The constraints that cannot be captured graphically are defined in OCL, as we illustrate in Fig.~\ref{fig:example}(b).
 Constraint \emph{C1} states that the age range for individuals is within the range $[0..100]$.
\emph{C2} states that the disability rate is within the range $(0..1]$ when the individual is disabled, and zero otherwise. 
\emph{C3} states that taxpayers who have an address in Luxembourg are considered residents. 
 \emph{C4} states that taxpayers with a local income but no local address are considered non-residents.
\emph{C5} states that a tax card applies only to  employment and pension incomes.

The constraint solving solution we propose in this article is aimed at generating \emph{valid} instantiations of CDs, meaning that the generated instance models should satisfy both the diagrammatic and OCL constraints.

\section{Background}\label{sec:background}
In this section, we provide background on OCL, search-based OCL solving, and SMT. 

\sectopic{OCL.} 
OCL is the de-facto constraint and query language for UML~\cite{OCL}. 
One can use OCL for defining: (1) invariants (conditions that must hold over all instances of a given class), e.g., \emph{C1} to \emph{C5} in Fig.~\ref{fig:example}(b), 
(2)~(user-defined) operations, 
e.g., our running example in Fig.~\ref{fig:example}(a) includes only one user-defined operation, named \textit{getAge}, whereas the remaining operations highlighted in blue in Fig.~\ref{fig:example}(b) are part of OCL, 
(3)~derived attributes (attributes whose values are calculated from other elements),
(4)~guard conditions for transitions in behavioral diagrams such as UML State Machines, and
(5)~pre- and post-conditions for user-defined operations.

We concentrate on handling \emph{OCL invariants} and \emph{the user-defined operations} needed by them. In our context, user-defined operations provide a convenient mechanism for factoring out queries which are used multiple times, or which are too complex to spell out within the invariants. Derived attributes are treated as user-defined operations. For example, the \textit{getAge} operation in Fig.~\ref{fig:example}(b) would be interchangeable with a derived attribute that calculates \hbox{the age of an individual (not shown).}

Guard conditions fall outside the scope of our work, since we deal exclusively with CDs and their instantiations.
Pre- and post-conditions are mainly an apparatus for verifying the executability and determinism of system behaviors~\cite{CabotCR09}. Although uncommon, pre- and post-conditions can also be used for constraining the instantiations of CDs. Despite such use of pre- and post-conditions being plausible, we leave pre- and post-conditions out of the scope of our work, and assume that all the relevant data restrictions are provided as invariants.

\sectopic{OCL constraint solving via (metaheuristic) search.} 
Search has been employed for addressing a wide range of software engineering problems~\cite{HarmanMZ12}.
Related to our objectives in this article, Ali et al.~\cite{Shaukat:2013,Shaukat:2016} develop a search-based method for solving OCL constraints.
In our previous work, we enhanced this method with new heuristics aimed at improving the exploration of the search space~\cite{Soltana:2017}.
The work we present in this article builds on our enhanced implementation of search-based OCL solving.
 This enhanced implementation is a contribution of our previous work~\cite{SoltanaTool} and not that of this article. Nevertheless, since the implementation is an intrinsic part of our approach, we outline its main technical characteristics below.

Our search-based OCL solver is based on the Alternating Variable Method (AVM)~\cite{AVMNew,AVAMF}. 
This choice is motivated by the empirical findings of Ali et al.~\cite{Shaukat:2013}, indicating that, for OCL constraint solving, AVM outperforms popular metaheuristic techniques such as (1+1) evolutionary~\cite{OnePlusOne} and genetic~\cite{Genetic} algorithms. AVM works in a similar manner to the well-known Hill Climbing algorithm~\cite{Hill}. In particular, it starts with an arbitrary solution; it then attempts to find a better solution by making incremental changes, and keeps evolving the solution until no further improvements can be found.

Concretely speaking and in our context, AVM starts with a random and potentially invalid instance model, $I$, which the algorithm later evolves so as to get it to satisfy the OCL constraints at hand. 
Specifically, we use AVM to optimize -- based on a fitness function that we describe momentarily -- a vector of variables $V = (v_{1},\dots, v_{n})$, where $n$ is the number of features in $I$ that should be evolved by search. For solving OCL constraints, the features are the objects (class instances), the association ends, and the attributes. For example, the \emph{disabilityRate} attribute of a specific instance of the \textit{PhysicalPerson} class would constitute a variable within $V$. We note that since AVM evolves the structure of $I$, e.g., by adding new objects, the length of $V$ is not fixed and may vary as the AVM algorithm unfolds. For example, if  $I$ includes no instance from  the \textit{PhysicalPerson} class, then there would be no variables representing the attributes and association ends of  \textit{PhysicalPerson} instances. However, if $I$ does include some instance of \textit{PhysicalPerson}, each attribute and each association end of that instance would give rise to one variable within $V$.
 
Given a non-empty $V$,  AVM optimizes the variables in $V$ as follows. It picks in a sequential manner a single variable $v_{i}$ from $V$. The algorithm then subjects $v_{i}$ to a search process called Iterated Pattern Search (IPS)~\cite{AVMNew}.
For a given $v_{i}$, IPS first makes small exploratory moves over $v_{i}$. For example, if $v_i$ is an integer, a small exploratory move could be incrementing or decrementing the value of $v_i$ by one unit. These exploratory moves are aimed at guessing what ``direction'' is more promising to explore, when the notion of direction is applicable. To illustrate, let us assume that $v_i$ is an integer with a current value of 20. IPS's first exploratory move could be to set $v_\text{i}$ to 21. If this increase leads to no improvement in the value of the fitness function, then IPS applies another exploratory move, whereby $v_\text{i}$ is decremented, e.g., to 19. Similar exploratory moves are defined for non-nominal variable types other than integers. For example, variables representing sets of objects are explored by adding or deleting an object. If no exploratory move leads to an improvement in the value of the fitness function, then no ``direction'' is worth further exploration. In this case, IPS terminates and AVM moves on to the next variable in $V$.
 
If one of the exploratory moves leads to an improvement in the fitness value, then a positive (increment) or negative (decrement) ``direction'' is established for making further moves. Based on the selected direction, moves with larger amplitudes will be made. For example, in the case of integers, increases or decreases by more than one unit will be applied. Moves in a certain direction will continue as long as the value of the fitness function keeps improving. When a given move deteriorates the fitness function value, the search has likely overshot an optimum. When this occurs, IPS goes back to the exploratory phase in order to re-establish a new direction for further optimizing the value of the variable in question.

For nominal variable types, e.g., enumerations, where the notion of direction does not apply, there is no IPS. In such cases, the variable in question is optimized as follows: AVM picks either (1) all the alternatives, when the space of all alternatives is small, in our case, containing $\leq 10$ alternatives, or (2) a random selection of the alternatives (in our case, 10 alternatives), otherwise. The alternative leading to the best fitness value is retained, and \hbox{AVM moves on to the next variable in $V$.}

Once all the variables in $V$ have been treated, AVM is said to have completed a ``search iteration.'' After the completion of a search iteration, AVM goes back to the first variable in $V$. AVM continues in this manner until $I$ becomes valid or a maximum user-specified number of search iterations has been reached. To reduce the chances of entrapment in non-solution regions of the search space, AVM relies on a simple heuristic: If a given search iteration yields no improvement at all in the fitness function value and $I$ is still invalid, then AVM abandons $I$ and continues with a fresh randomly created instance model $I'$.

Our implementation of search-based OCL solving employs the fitness function defined by Ali et al.~\cite{Shaukat:2013}. Essentially, this function is a quantitative assessment of how far a given instance model is from satisfying a given set of OCL constraints. For details, we refer the reader to Ali et al. To facilitate comprehension, we briefly illustrate the fitness function using a simple example. Suppose that the set of constraints to satisfy includes only the following constraint, $C$:
\ocl{TaxPayer.\makeBlue{allInstances}()->\makeBlue{forAll} (x\ourMid x.incomes->\makeBlue{size}()>=1 \makePurple{and} x.incomes->\makeBlue{size}()<4)}. Further, suppose that the instance model being evolved has two instances of the \textit{TaxPayer} class, one with a single income and the other with four incomes. 
The computed distance for the first taxpayer is 0, since that taxpayer already satisfies $C$; whereas the distance for the second taxpayer is 1, since this second taxpayer has one more income than what $C$ allows. The distance for \makeBlack{\ocl{forAll}} and thus for $C$ as a whole is computed as the average of the distances for all (here, the two) taxpayers: $(0 + 1)/2 = 0.5$. Now, let $C'$ be the same as $C$, but with $C$'s \ocl{\makeBlack{forAll}} operation replaced with \ocl{\makeBlack{exists}}. For the same instance model, the fitness function for $C'$ would be computed as $min\,(0,1) = 0$. An overall distance of 0 indicates that a solution has been found; otherwise, AVM will iteratively attempt to minimize \hbox{the (overall) distance, as described above.}

\sectopic{SMT.} 
SMT is the problem of determining whether a constraint expressed in first-order logic is satisfiable, when certain symbols have specific interpretations in a combination of background theories~\cite{MouraB11}. 
State-of-the-art SMT solvers, e.g., Z3~\cite{z3} and CVC4~\cite{CVC4}, accept as input a standard language, called SMT-LIB~\cite{SMT-LIB}.
For example,  \emph{C2} of Fig.~\ref{fig:example}(b) can be expressed in SMT-LIB as follows:\,\ocl{(\makePurple{ite}}\,\ocl{(= x None)}\,\ocl{(= y 0)}\,\ocl{(\makePurple{and}}\,\ocl{(> y  0)}\,\ocl{(<= y  1))))}, where \ocl{\makeBlack{x}} and \ocl{\makeBlack{y}} respectively represent the disability type and the disability rate for a given person.

Most SMT solvers employ a SAT solver alongside the decision procedures for the background theories they support~\cite{KatzBTRH16}.
The SAT solver operates on Boolean formulas, and the decision procedures operate on the symbols within the formulas. 
SMT solvers are highly efficient for solving quantifier-free constraints, when suitable background theories exist. 
Nevertheless, when constructs such as quantification, collections, or symbols with no background theories are added to the mix, these solvers often become inefficient.

Since SMT solvers alone are not able to efficiently handle all types of OCL constraints that are commonly used for specifying software systems, we propose to exploit SMT only for solving OCL subformulas where SMT is very likely to deliver efficiency gains. 

\section{Approach}\label{sec:approach}
\begin{figure}[h]
	\centering
	\includegraphics[width=0.8\linewidth]{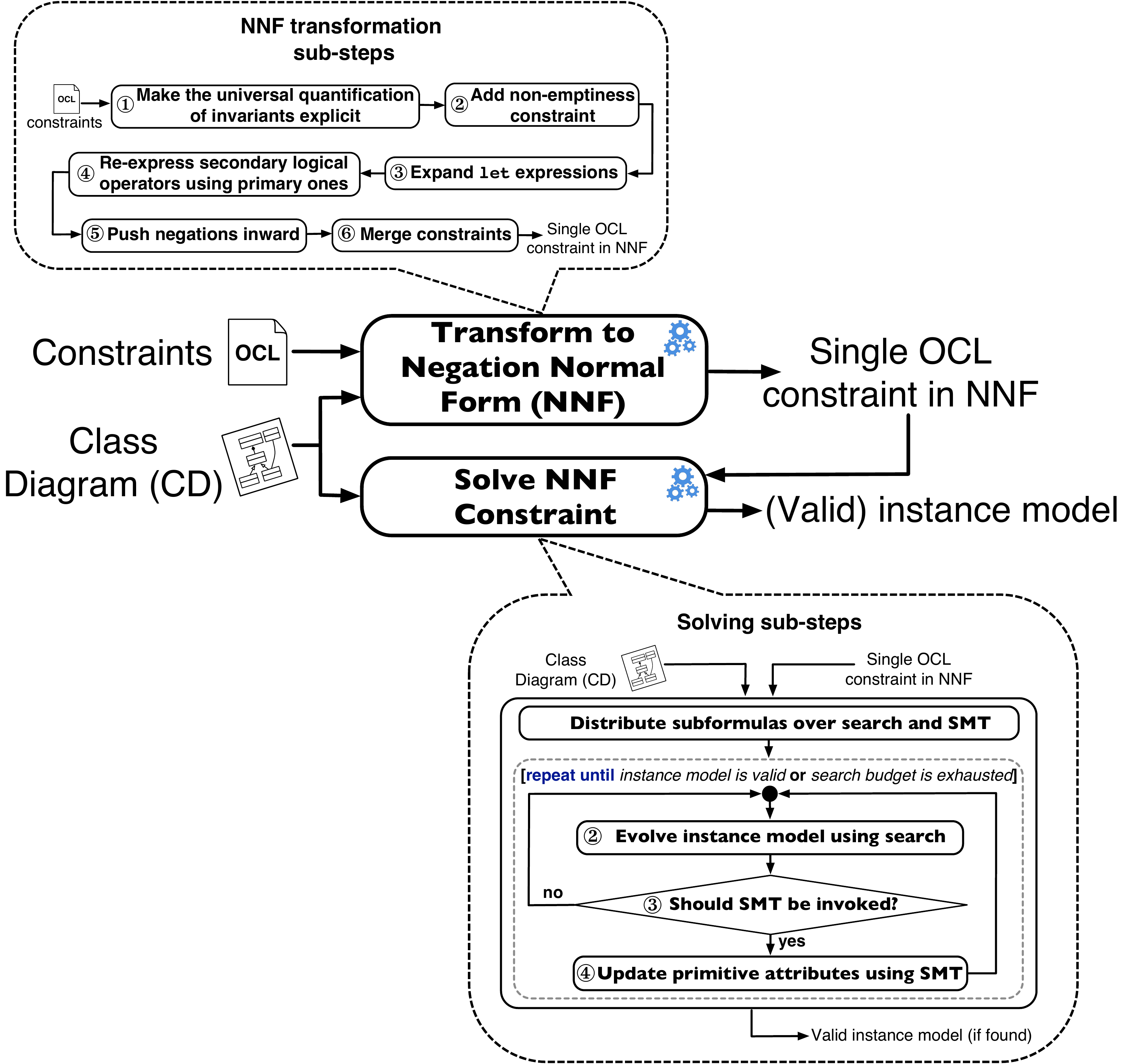}
	\caption{Approach Overview\label{fig:approach}}
\end{figure}

Fig.~\ref{fig:approach} presents an overview of our approach, hereafter referred to as PLEDGE (PracticaL and Efficient Data GEnerator for UML), i.e., the name of the tool implementing it as mentioned in the introduction. 
PLEDGE takes as input a CD and a set of OCL constraints. 
The output is a valid instance model if one can be found.

PLEDGE works in two phases: First, it constructs a single OCL constraint in Negation Normal Form (NNF) that is logically equivalent to the conjunction of the constraints that need to be solved. A constraint is in NNF if it uses only the primary Boolean operators, i.e., conjunction, disjunction and negation, and further has all the negations pushed inward to the atomic level~\cite{NNFBook}. The subs-steps underlying the NNF transformation are further discussed in Section~\ref{sec:ocl}. We use NNF because it facilitates defining and distributing the solving tasks over search and SMT, while also simplifying the translation from OCL to SMT-LIB. Specifically, the single constraint is a conjunction of the following: (1) the input OCL constraints, and (2) OCL representations of the multiplicity constraints in the input CD. For example, the \ocl{1..*} cardinality attached to the target end of the \textit{earns} association in Fig.~\ref{fig:example}(a) is represented in OCL as: \ocl{\makeBlue{self}.incomes->\makeBlue{size}() >= 1}, where \makeBlack{\ocl{self}} refers to any instance of \textit{TaxPayer}. 
We leave out of this conjunction the basic constraints imposed by the CD's diagrammatic notation, e.g., type conformance for the association ends and the non-instantiability of abstract classes, as illustrated in Section~\ref{sec:example}. PLEDGE implicitly enforces these basic constraints when creating or tweaking instance models.

The second phase of PLEDGE is to solve the resulting NNF constraint. 
To do so, we utilize a combination of search and SMT, with the solving tasks distributed in the following manner:  We have search handle subformulas whose satisfaction involves structural tweaks to the instance model, i.e., additions and deletions of objects and links. An example constraint (subformula) handled by search is \emph{C5} of Fig.~\ref{fig:example}(b). If the instance model happens to violate \emph{C5}, any successful attempt at satisfying the constraint will necessarily involve adding / removing objects and links.

We have SMT handle subformulas which exclusively constrain attributes with primitive types. Many such subformulas, e.g., those 
whose symbols are within linear arithmetic, can be efficiently handled by background SMT theories. For example, we use SMT to assign a value to the \textit{birthYear} attribute of an instance of \textit{PhysicalPerson} in such a way that \emph{C1} of Fig.~\ref{fig:example}(b) is satisfied. Finally, we have \emph{both} search and SMT handle subformulas whose satisfaction may require a combination of structural tweaks and value assignments to primitive attributes. For example,  satisfying \emph{C3} of Fig.~\ref{fig:example}(b) may involve both adding instances of \textit{Address} and setting the \textit{country} and \textit{isResident} attributes of \textit{Address} and \textit{TaxPayer} instances.

As mentioned in the introduction, the core idea behind  PLEDGE is to enhance the performance of search-based OCL constraint solving by combining it with SMT.  Specifically, PLEDGE aims to support search with SMT solving capabilities for the subformulas where efficient and effective decision procedures for certain background theories exist. The decision procedures that PLEDGE taps into are those related to theories for basic Boolean operators, and arithmetic over integer, real, and floating-point numbers~\cite{dechter2003constraint,SMT}. This is ensured by applying SMT over only the subformulas from the NNF constraint that  concern attributes with primitive types. All remaining subformulas, which are mostly concerned with constraints on objects and links, are delegated to search. As illustrated earlier, this type of subformulas requires  evolving the structure of the instance model to be satisfied. Further, these subformulas often involve, among other things, quantifications, collection operations, and type operations,  which are not currently (well) supported by SMT (see Section~\ref{sec:background}).

Nevertheless, one should \emph{not} view the distribution of the solving tasks as a clear-cut division of the NNF constraint into smaller constraints that can be later solved independently. 
For example, and as illustrated earlier,  SMT and search might have overlapping solving tasks where the involved subformulas require a combination of structural tweaks and value assignments to primitive attributes.  Further, in the PLEDGE approach,  SMT only operates over the primitive attributes of the objects already created via search (fixed universe). For example, satisfying \emph{C1} and  \emph{C2}  of Fig.~\ref{fig:example}(b), which are fully assigned to SMT, is only warranted  for the objects of type \textit{PhysicalPerson} that were created via search, e.g., created to satisfy multiplicity or other user constraints. As a result, \emph{C1} and  \emph{C2}  cannot be solved independently by SMT regardless of how search evolves the instance model.  Rather and as discussed in Section~\ref{sec:solving}, search and SMT modify a shared instance model in order to satisfy the NNF constraint, with each technology focusing on the subformulas delegated to it. We further discuss in more detail how we delegate the solving tasks in Section~\ref{sec:share}.

We note that the generated data is stored as a UML instance model and thus can be used directly for model-based testing in a UML context~\cite{Baudry:2009,KessentiniSB11}. If a different test input format is required, say, text files, databases or a combination thereof, the generated instance model needs to be adapted. For simple CDs that do not involve inheritance or association classes, the objects and links in an instance model can be converted relatively easily into other representations. In particular, one can employ model-to-text or model-to-model transformation using tools such as Acceleo~\cite{Acceleo}, ATL~\cite{ATL} and Xpand~\cite{Xpand} for conversion into the desired test-input format. In an object-oriented context and when CDs have inheritance or association classes, it is more practical to use an object-relational mapping (ORM) tool for adapting the test input~\cite{TorresGPM17}. To this end, one may use tools such as Hibernate~\cite{Hibernate} or Apache OpenJPA~\cite{OpenJPA}. In the remainder of the article, we take an instance model to be equivalent with test data.

We further note that our examples above refer directly to the original constraints in Fig.~\ref{fig:example}, rather than to the corresponding fragments in the derived NNF constraint. This is only to ease illustration; the whole solving process, including making decisions about which subformulas to delegate to search, SMT or both, is with respect to the NNF constraint resulting from the first phase of PLEDGE. All the technical subs-steps underlying the solving phase are discussed in depth in Section~\ref{sec:solving}.

\section{Transformation to Negation Normal Form}\label{sec:ocl}
\begin{figure}[h]
	\centering
	\includegraphics[width=0.8\linewidth]{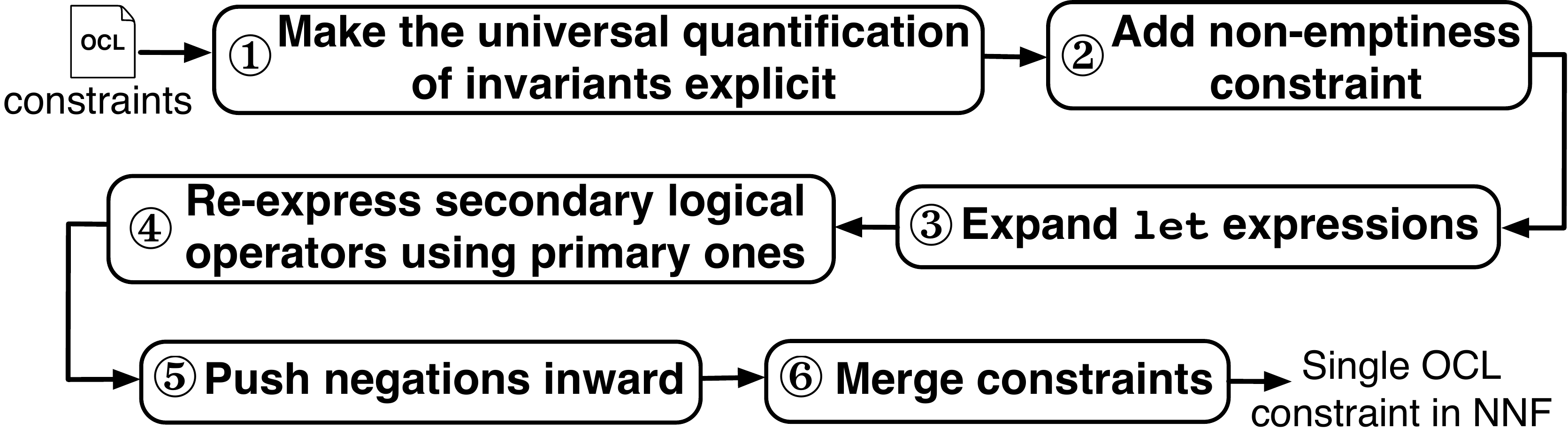}
	\caption{Pipeline for Deriving the NNF Constraint to Solve\label{fig:pipeline}}
\end{figure}

In this section, we elaborate the first phase of the PLEDGE approach outlined in Section~\ref{sec:approach}. 
The transformation to NNF is performed using the pipeline of Fig.~\ref{fig:pipeline}. 

\begin{figure}[b]
	\centering
	\includegraphics[width=0.8\linewidth]{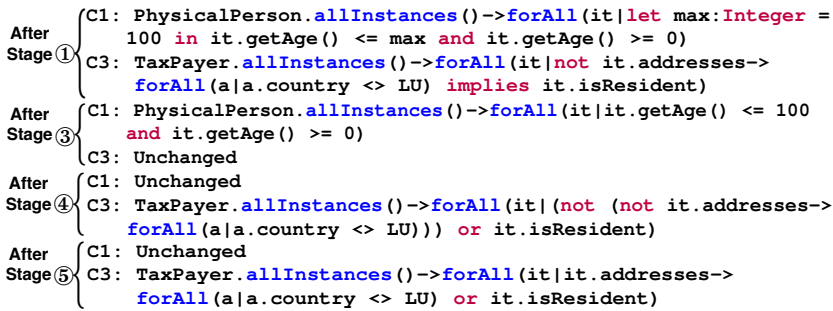}
	\caption{{Illustration of NNF Transformation over \emph{C1} and \emph{C3} of Fig.~\ref{fig:example}(b)
	\label{fig:transExample}
	}}
\end{figure}

As noted in Section~\ref{sec:background}, our focus is on constraints expressing OCL invariants. Invariants are implicitly universally quantified. In Step~1 of the pipeline of Fig.~\ref{fig:pipeline}, we make this implicit quantification explicit. Doing so enables us to later merge different constraints (Step~6), even when the constraints do not share the same context.

In Fig.~\ref{fig:transExample}, we show the result of applying Step~1 to \emph{C1} and \emph{C3} of Fig.~\ref{fig:example}(b).
For efficiency reasons, Step~1 binds only one universal quantifier to a given OCL context. This is achieved by taking the conjunction of all the invariants that have the same context, before universal quantification is made explicit. For example,  \emph{C1} and  \emph{C2} of Fig.~\ref{fig:example}(b) are both defined over \textit{PhysicalPerson}. Step~1 would thus consider \emph{C1} \ocl{and} \emph{C2} as one constraint. Due to space, we do not illustrate this treatment in  Fig.~\ref{fig:transExample}.

Step~2 adds a user-defined non-emptiness constraint to the set of constraints resulting from Step~1.
The non-emptiness constraint excludes the empty solution and thus avoids vacuous truth, noting that the constraints from Step~1 are invariants and thus always satisfied by the empty instance model. The non-emptiness constraint is, in contrast, not universally quantified. In the simplest case, this constraint states that a certain class of the input CD must have an instance. A natural non-emptiness constraint for our running example would be \hbox{\ocl{TaxPayer.\makeBlue{allInstances}()->\makeBlue{size}() >= 1}}, stating that the instance model must contain at least one instance of \textit{TaxPayer}. 
In practice,  the non-emptiness constraint can be more elaborate, e.g., bounding the maximum number of instances to generate from each class.   Regardless of its complexity, the non-emptiness constraint must always be violated by the empty instance model and its satisfaction has to involve the creation of certain objects within the instance model.

Step~3 of the pipeline performs an inline expansion of any \ocl{\makeBlack{let}} expressions within the constraints. These expressions are commonly used for avoiding repetition and improving readability. For our OCL solving process to be sound, we need to expand all \ocl{\makeBlack{let}} expressions, including nested ones. 
Fig.~\ref{fig:transExample} illustrates how the \ocl{let} expression in \emph{C1} is expanded inline.

Step~4 re-expresses OCL's secondary Boolean operators, i.e., \ocl{\makeBlack{implies}}, \ocl{\makeBlack{xor}} and \ocl{\makeBlack{if-then-else}}, in terms of the primary ones, i.e., \ocl{\makeBlack{and}}, \ocl{\makeBlack{or}} and \ocl{\makeBlack{not}}. 
Step~5 pushes the negation operators inward by applying De Morgan's law~\cite{Hurley:2014}. 
Both Steps 4~and~5 are illustrated in Fig.~\ref{fig:transExample} over \emph{C3}.

Steps~4 and 5 are the core steps for transforming the initial OCL constraints to NNF. These steps always terminate. In particular, Step~4 terminates when the OCL constraints become free of OCL's secondary Boolean operators; Step~5 terminates when no negation operator can be pushed further inward. The logical rules underlying these steps are equivalence preserving~\cite{Hurley:2014}, meaning that the rules transform any constraint $C$ into an NNF constraint $C'$ such as $C$ and $C'$ share the same truth table. In terms of scalability, the main risk associated with the NNF transformation is that Step~4 can lead to an exponential blowup in the size of the NNF constraint. 
Despite such a blowup being theoretically possible, the situation is unlikely to pose issues in practice. This is because the transformation is applied to OCL constraints that have been either manually written by analysts or derived from CDs, as explained in Section~\ref{sec:approach}. In the former case, individual constraints would typically include at most only a handful of secondary Boolean operators,  considering that a large number of such operators would considerably increase cognitive load and decrease understandability. In the latter case, the derived constraints are free of secondary Boolean operators and thus not affected by Step~4. In  all of our case  studies reported in Section~\ref{sec:evaluation}, the  NNF transformation took negligible time (in the order of milliseconds).

\begin{figure}[b]
	\centering
	\includegraphics[width=0.85\linewidth]{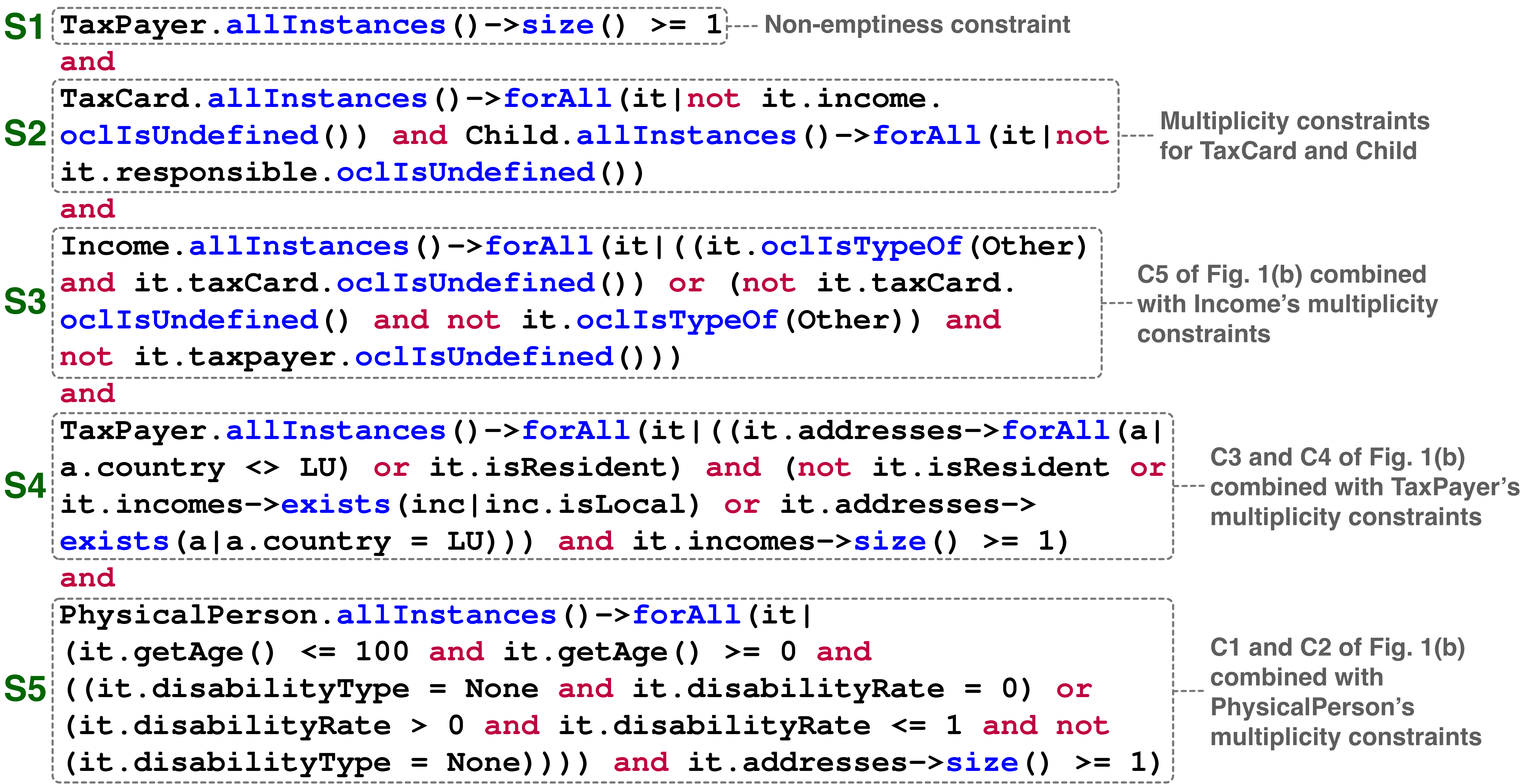}
	\caption{Example NNF Constraint Produced by the Pipeline of Fig.~\ref{fig:pipeline} \label{fig:NNF}}
\end{figure}

Finally, Step~6 takes the conjunction of the constraints obtained from Step~5. The resulting NNF constraint for our running example is shown in Fig.~\ref{fig:NNF}. For easier reference, we divide this constraint into five subformulas, labeled S1 to S5.

We need to point out two subtleties about the pipeline of Fig.~\ref{fig:pipeline}: First, Steps 1 and 3 have to be done in such a way that no name clashes arise. To avoid name clashes, we give all variables distinct (automatically generated) names. Since no name clashes occur in our running example, we elected, for readability reasons, not to rename the variables in Figs.~\ref{fig:transExample}~and~\ref{fig:NNF}. Second, Steps~3 through~5 of the pipeline further apply to the user-defined operations needed by the constraints. In our example, the only (user-defined) operation needed, \textit{getAge}, remains unaltered since it has no \ocl{let} expressions, secondary logical operators, or negations.

The NNF constraint generated by the pipeline of Fig.~\ref{fig:pipeline} is solved through the process that we present next.

\section{Hybrid OCL Solving Using Search and SMT}\label{sec:solving}
\begin{figure}[h]
	\centering
	\includegraphics[width=0.6\linewidth]{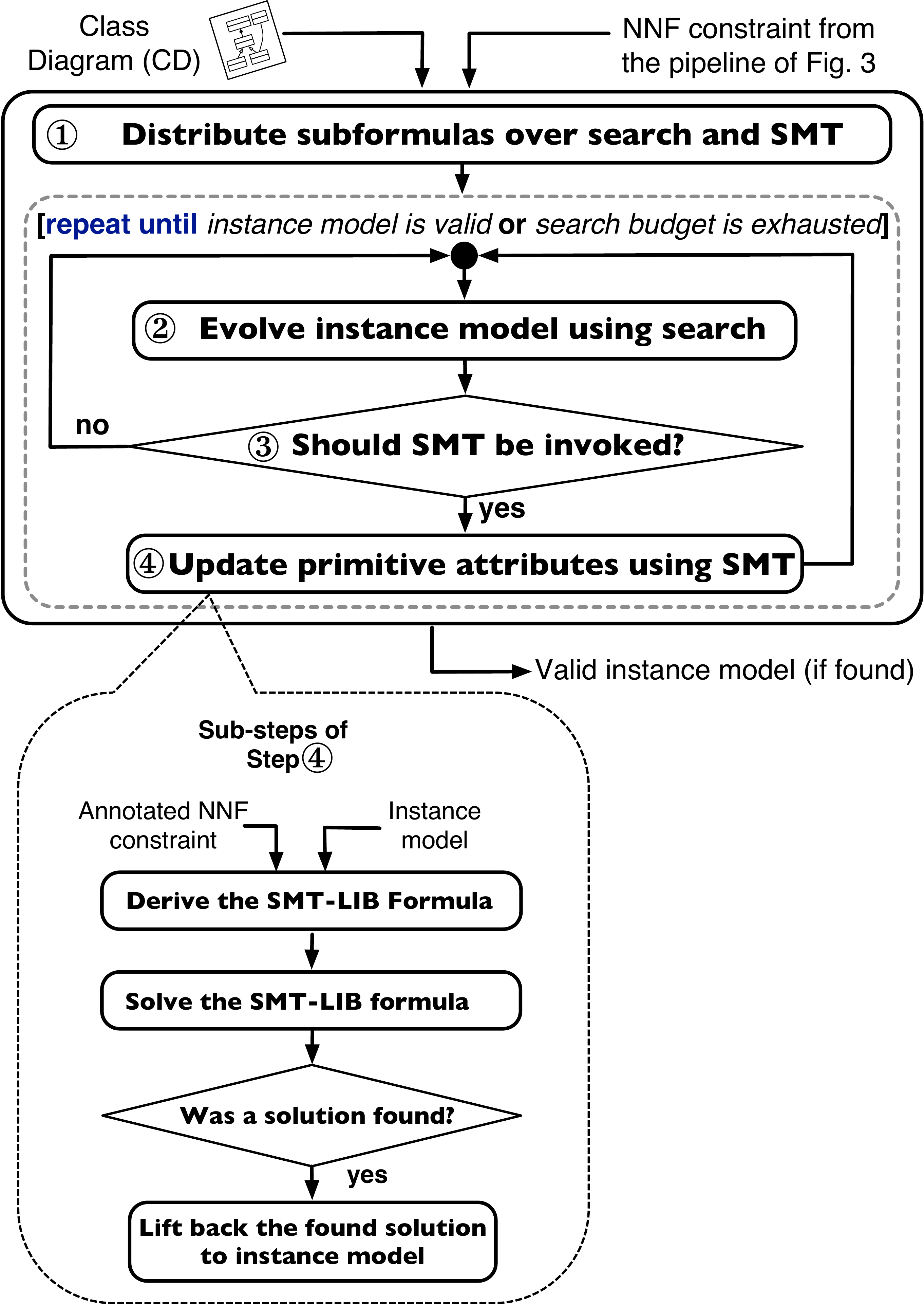}
	\caption{Hybrid OCL Solving Process \label{fig:process}}
\end{figure}

In this section, we elaborate our hybrid OCL solving process, i.e., the second phase of the PLEDGE approach in Section~\ref{sec:approach}.
The solving process, depicted in Fig.~\ref{fig:process}, has four steps. The first step is performed once; the remaining three steps are iterative. 
These three steps are repeated until either an instance model satisfying the NNF constraint is found, or the maximum number of search iterations is reached. 
We discuss each step of the solving process below.

\subsection{{Delegating Solving Tasks to Search \& SMT (Step\,1,\,Fig.\,\ref{fig:process})}}\label{sec:share}

Step~1 of the process of Fig.~\ref{fig:process} decides how to delegate to search and SMT the solving of different subformulas of the NNF constraint derived in Section~\ref{sec:ocl}.
Specifically, Step~1 applies a static and deterministic procedure, whereby the nodes in the Abstract Syntax Tree (AST) of the derived NNF constraint are given one of the following labels: (1)~\textit{search}, (2)~\textit{SMT}, or (3)~\textit{both}.

These labels operationalize our strategy, described in Section~\ref{sec:approach}, for combining search and SMT. Before presenting our labeling procedure in detail, we intuitively describe how it works: The procedure attaches labels to the AST nodes of an NNF constraint based on each node's type, properties, and position within the AST. For example, S5 in Fig.~\ref{fig:NNF} has two subformulas of the form \ocl{x\,>=\,$\langle$constant$\rangle$}, where \ocl{x} is a variable. Although the AST root nodes for both subformulas are the \ocl{>=} operation, these root nodes are labeled differently. Specifically, the root node of the first subformula, \ocl{it.getAge() >= 0}, is labeled \textit{SMT}; this is because the inequality here is indirectly defined over the \textit{birthYear} primitive attribute in Fig.~\ref{fig:example}(a). In contrast, the root node of the second subformula,  \ocl{it.addresses-> \makeBlue{size}() >= 1},  is labeled \textit{search}, because the inequality here restricts the minimum number of allowed instantiations of the \textit{resides at} association in Fig.~\ref{fig:example}(a). We recall from Section~4 that we want to have search handle all subformulas that are exclusively concerned with the structure of the instance model. For the sake of argument, let us hypothetically assume that we have a third subformula, which is the conjunction of the two subformulas mentioned above.
In such a case, the AST root node representing the conjunction would be labeled \textit{both}. This because solving this (hypothetical) subformula would involve both search and SMT.

More precisely and technically speaking, the labeling procedure of Step~1 is realized by a depth-first traversal of the AST of a given NNF constraint, with the visited nodes labeled via a set of predefined rules. The rules are heuristics that we have designed to assign, in a deterministic manner, a single label to each visited AST node, except those representing constants (e.g., 100 in Fig. 7, discussed later); constants do not play a role in deciding about the delegation of solving tasks. Our rules cover all the AST node types defined by the OCL 2.4 metamodel~\cite{OCL}. Since the rules are numerous, we do not list them here; our complete rule set and the rationale behind each rule can be found in Appendix~\ref{sec:rules}.
Below, we provide a general description of the rules and illustrate them over S5 of the NNF constraint in Fig.~\ref{fig:NNF}.

\begin{figure}[t]
	\centering
	\includegraphics[width=0.8\linewidth]{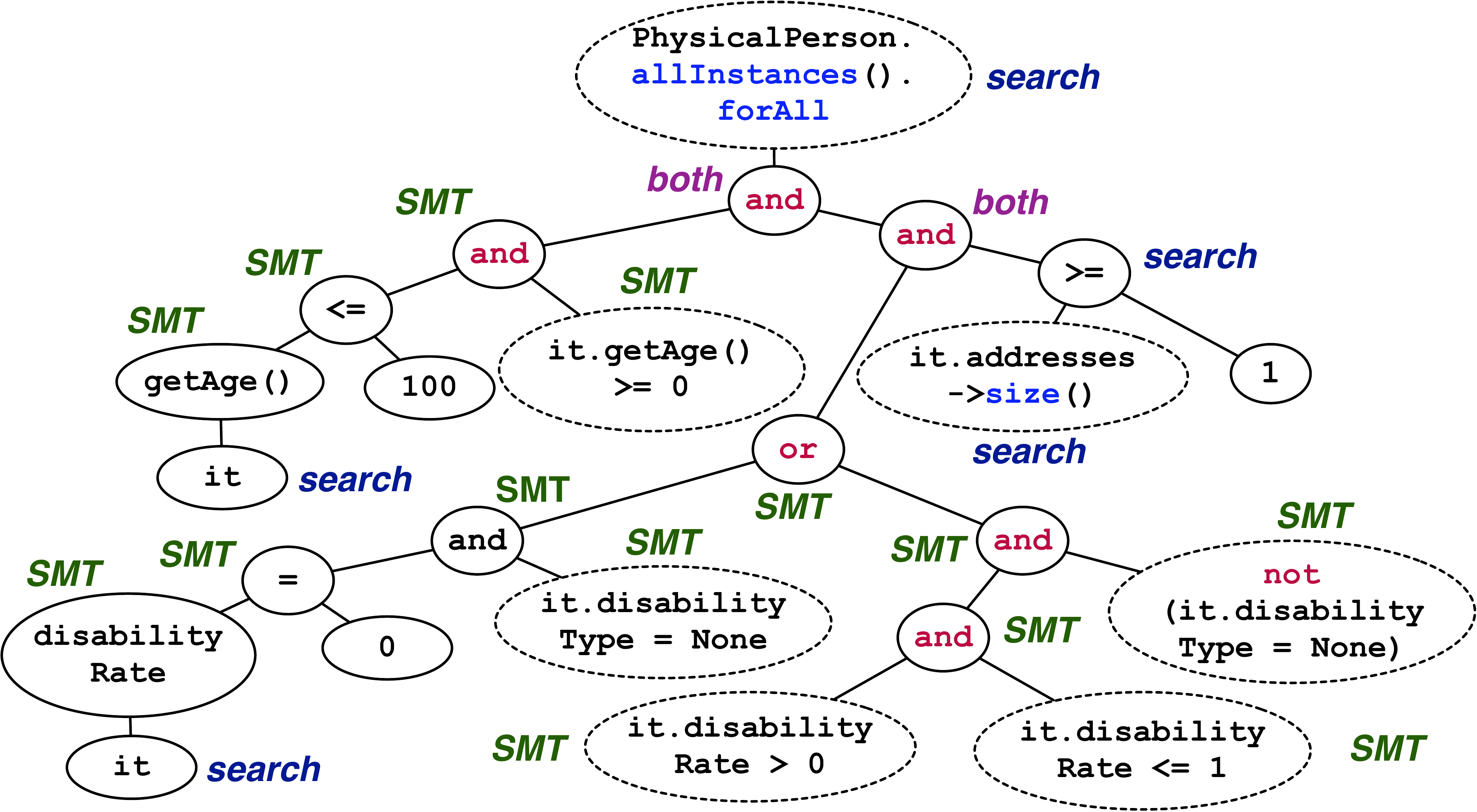}
	\caption{Abstract Syntax Tree for S5 of the Constraint in Fig.~\ref{fig:NNF} \label{fig:intermediate}}
\end{figure}

The AST for S5 is shown in Fig.~\ref{fig:intermediate}. To avoid clutter, we have collapsed some nodes. The collapsed nodes, marked by dashed borders, expand similarly to the node with the ``\ocl{=}'' symbol (bottom left of the figure). The label shown for each collapsed node is that of its root. Our rules can be broadly classified into three categories. The rules in the first category label an AST node based only on information contained within the node. Notably, (i)~nodes representing quantifiers or object references are labeled \textit{search}, e.g., the root node of the AST of Fig.~\ref{fig:intermediate}; and (ii)~nodes representing calls to user-defined operations are labeled \textit{SMT} if  the operations are not recursive and have primitive return types (e.g., the \ocl{\makeBlack{getAge}} nodes in Fig.~\ref{fig:intermediate}); otherwise, these nodes are labeled \textit{search}.

The rules in the second category infer a label for an AST node based on the type of the node itself and those of its ancestors. Notably, a node representing a primitive attribute is labeled \textit{SMT}, unless the node has an ancestor of a certain type, e.g., \ocl{exists}, in which case, the attribute is labeled \emph{both}. In Fig.~\ref{fig:intermediate}, all primitive attributes are labeled \textit{SMT}. To provide an example of a primitive attribute that is labeled \emph{both}, consider S4 in the constraint of Fig.~\ref{fig:NNF}. In the AST for S4 (not shown), the node representing \textit{isLocal} would be labeled \textit{both}. 

The rules in the third category infer a label for an AST node based on the node's children.
Notable constructs handled by this category of rules are logical and numerical operators.
For example, the \ocl{and}, \ocl{or}, \ocl{not}, \ocl{=}, \ocl{>=}, \ocl{<=}, \ocl{>} nodes in Fig.~\ref{fig:intermediate} are all labeled based on their (immediate) children. Specifically, if all children have the same label, that label is propagated to the parent node. 
If the children have different labels, then the parent node is labeled \textit{both}. AST nodes representing constants are left unlabeled; these nodes do not play a role in deciding about the delegation of solving tasks.

Although not illustrated in Fig.~\ref{fig:intermediate}, when a node representing a call to a user-defined operation is labeled \textit{SMT}, the AST of the operation is subject to the same labeling procedure discussed above. This treatment does not apply when these nodes are labeled \textit{search}; in such cases, all the nodes within the AST of the referenced operation are labeled \textit{search}. This is because, as we elaborate in Section~\ref{sec:z3}, the PLEDGE approach hides from SMT any user-defined operation called from a node that is labeled \textit{search}. Consequently, we need to handle any such operation call via search only.

Noting that OCL has its roots in first-order logic, we believe that our delegation rules are likely to be adaptable to other formal specification languages based on first-order logic. At the same time, we need to stress that our current rules have been defined directly over the OCL metamodel. Assessing whether and how these rules would generalize beyond OCL requires additional investigations, which we leave for future work.

\subsection{The Search Step (Step~2, Fig. \ref{fig:process})}\label{sec:search}
Step~2 runs one iteration of search.
This step utilizes the search-based OCL solver we developed previously \cite{Soltana:2017} and outlined in Section~\ref{sec:background}.
We refer to this solver as the \emph{baseline} hereafter.
We modify the baseline by limiting what search is allowed to do. Precisely, we allow search to manipulate only element types that are referenced by AST nodes labeled \textit{search} or \textit{both}. For example, if we were to satisfy S5 of the constraint in Fig.~\ref{fig:NNF}, search would be allowed to add / remove instances of only the following types: the \textit{PhysicalPerson} and \textit{Address} classes and the \textit{resides\,at} association. Suppose that search creates an instance $P$ of \textit{PhysicalPerson}. In doing so, $P$'s primitive attributes necessarily receive initial values (e.g., default or random values). However, once $P$ is added to the instance model, search is prohibited from tweaking $P$'s primitive attributes. This is because the primitive attributes of \textit{PhysicalPerson} are referenced only within nodes that \hbox{are labeled \textit{SMT} in the AST of Fig.~\ref{fig:intermediate}.}

\subsection{Avoiding Futile SMT Invocations (Step 3, Fig.~\ref{fig:process})}\label{sec:check}
Step 3 is aimed at avoiding SMT invocations that, given the current structure of the instance model, have no chance of solving the NNF constraint. We recall from Section~\ref{sec:approach} that the only means we make available to SMT for altering an instance model is by manipulating primitive attribute values. If a violation is unfixable this way, invoking SMT will be futile. For example, if the current instance model violates some multiplicity constraint, the violation cannot be fixed by SMT in our proposed solving process.

\begin{figure}[h]
	\centering
\includegraphics[width=0.4\linewidth]{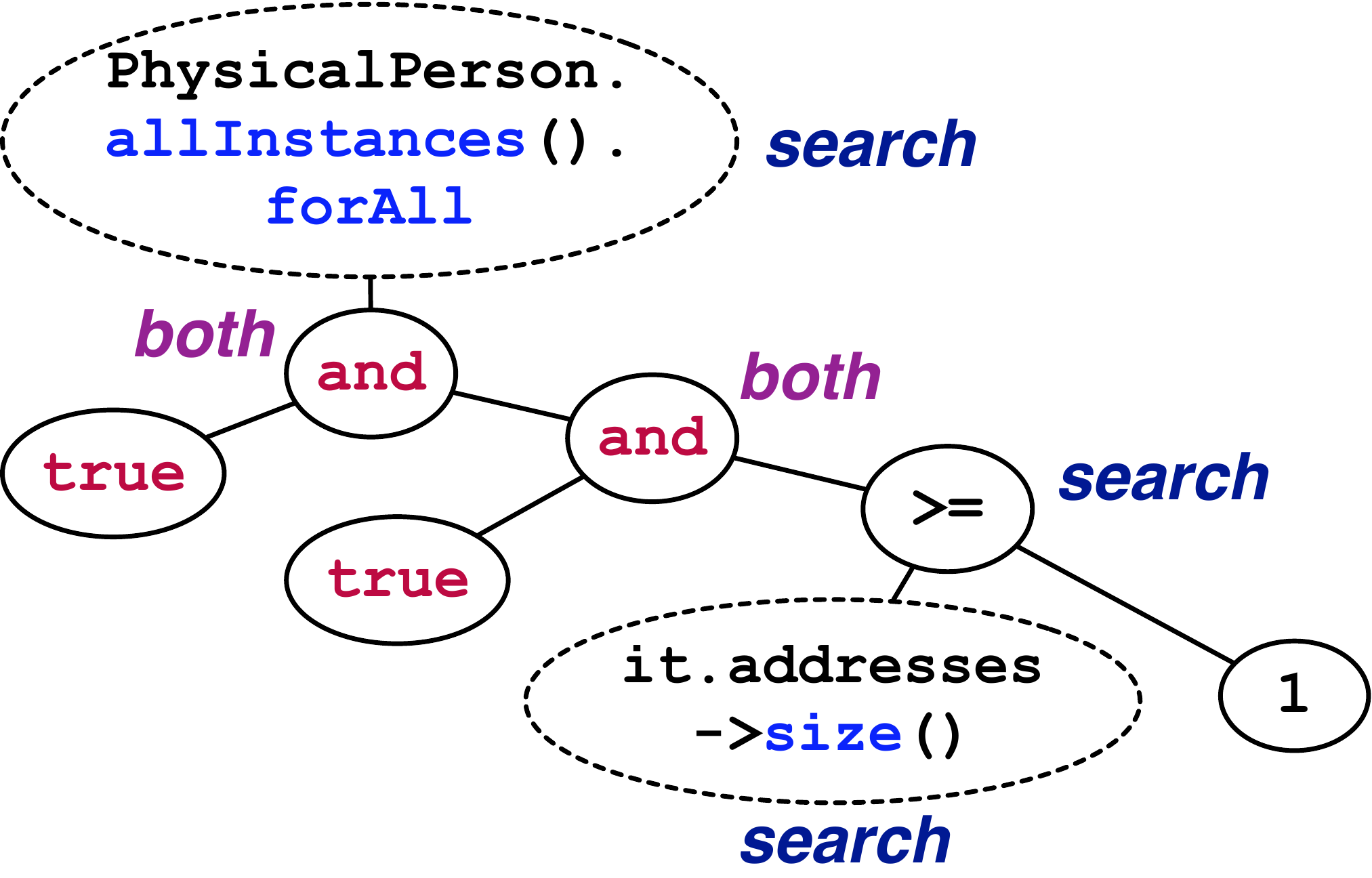}
\caption{Reduced AST obtained from the AST in Fig.~\ref{fig:intermediate} \label{fig:pruned}}
\end{figure}

To decide whether to invoke SMT, we do as follows: First, we clone the (labeled) AST of the underlying NNF constraint. Next, in this (cloned) AST, we replace with \ocl{true} any node labeled \textit{SMT} and representing a Boolean expression. Naturally, the branches below the replaced nodes are pruned.
In Fig.~\ref{fig:pruned}, we show the result of this treatment applied to the AST of Fig.~\ref{fig:intermediate}. 
The rationale behind the treatment is simple: We assume that SMT, \emph{if} invoked, will be able to solve all subformulas delegated to it. If the constraint represented by the reduced AST evaluates to \ocl{false} on the current instance model, SMT cannot conceivably satisfy the whole constraint. Otherwise, we give SMT a chance to solve the subformulas delegated to it.

We make two remarks about Step 3. First, for the step to be sound, the original AST must represent an OCL expression in NNF, where negations have been pushed all the way inward, as discussed in Section~\ref{sec:ocl}. Second, the construction of the reduced AST is a one-off activity. It is only the evaluation of this AST that needs to be repeated in each iteration of Step 3.

\subsection{The SMT Step (Step 4, Fig.~\ref{fig:process})}\label{sec:z3}

For a given instance model, Step~4 does the following: It first encodes into an SMT-LIB formula the solving task that needs to be handled by SMT. Next, it invokes an SMT solver on the resulting formula. The formula is solved if the SMT solver succeeds in finding a satisfying assignment to the SMT-LIB variables, where each variable corresponds to a specific primitive attribute in the instance model. If the formula is solved, the solution will be ``lifted back'' to the instance model.  By lifting back, we mean that the value of each SMT-LIB variable in the solution is assigned to its corresponding primitive attribute in the instance model. For example, suppose that an instance model passes the check in Step~3 (Section~\ref{sec:check}) but still violates S5 of the NNF constraint in Fig.~\ref{fig:NNF}. This means that the attributes in some instances of \textit{PhysicalPerson} (individuals) have to be re-assigned so that the instance model would satisfy S5.
In this case, S5 is violated because the instance model includes individuals who: (1)  have an age outside the range $[0..100]$, thus violating \emph{C1} in Fig.~\ref{fig:example}(b), (2) are disabled but have a disability rate outside the range (0..1], thus violating \emph{C2} in Fig.~\ref{fig:example}(b), or (3) are not disabled but have a non-zero disability rate, thus also violating \emph{C2} in Fig.~\ref{fig:example}(b). Here, the SMT solver would be tasked with finding appropriate values for the \emph{birthYear}, \emph{disabilityType} and \emph{disabilityRate} attributes of the \textit{PhysicalPerson} instances already in the instance model such that \emph{C1} and \emph{C2} in Fig.~\ref{fig:example}(b) would hold.

\begin{algo}[t]
\caption{Derive SMT-LIB Formula from NNF Constraint (\textbf{deriveFormula}) \label{algMain}}
\SetAlgoLined
\DontPrintSemicolon 
\SetInd{0.4em}{0.4em}
\SetKwInOut{Parameter}{Fun. calls}
 \SetKwInOut{Input}{Inputs}\SetKwInOut{Output}{Output}
 \setcounter{AlgoLine}{0}
 
\Input{
(1) The (labeled) AST, $T$, of an NNF constraint; (2) an instance model, \var{inst}.
}
\BlankLine 
\Output{An SMT-LIB formula over the primitive attributes of \var{inst} referenced in $T$.}
\BlankLine 
\Parameter{
\textbf{expand}: expands over an instance model the primitive-attribute-carrying quantifiers of an OCL expression represented by an AST  (Alg.~\ref{algExpand}); 
\textbf{substitute}: substitutes all subformulas delegated exclusively to search by their concrete evaluation over an instance model (Alg.~\ref{algSubstitute}); 
\textbf{processOperations}: performs special processing over the ASTs of user-defined operations so that these operations can later be translated to SMT-LIB (Alg.~\ref{algOperations});  
\textbf{fromOCL2SMT-LIB}: generates an SMT-LIB formula for a (processed) NNF constraint (Alg.~\ref{algToSMT-LIB}).
\BlankLine 
}
Let \var{root} be the root node of $T$\\ 
\var{root\textsubscript{expanded}} $\leftarrow$ \textbf{expand}(\var{root}, \var{inst})\\
\var{root\textsubscript{post-substitution}} $\leftarrow$ \textbf{substitute}(\var{root\textsubscript{expanded}}, \var{inst})\\
Let $\mathcal{U}$ be the set of the ASTs of all the user-defined operations referenced (directly or indirectly) within the constraint represented by $T$\\
$\mathcal{U}_{\textrm{processed}}$  $\leftarrow$ \textbf{processOperations}($\mathcal{U}$, \var{root\textsubscript{post-substitution}}, \var{inst})\\[.5em]
\Return{\rm\textbf{OCL2SMT-LIB}(\var{root\textsubscript{post-substitution}}, $\mathcal{U}_{\textrm{processed}}$, \var{inst})}\\
\end{algo}

\subsubsection{Deriving the SMT-LIB Formula} 
Alg.~\ref{algMain}, named \textbf{deriveFormula}, presents our algorithm for constructing an SMT-LIB formula. 
The algorithm takes the following as input: (1) the AST of an NNF constraint already processed by the labeling procedure of Section~\ref{sec:share}, and (2) an instance model passing the check discussed in Section~\ref{sec:check}. Initially, \textbf{deriveFormula} expands the primitive-attributes-carrying quantifiers of the NNF constraint over the instance model (L.~2). The expansion hides quantification, thus allowing SMT to be invoked on a quantifier-free formula -- this is where, as we noted in Section~\ref{sec:background}, existing SMT solvers are most efficient. After expansion, \textbf{deriveFormula} substitutes any subformula that is exclusive to search with the concrete evaluation of that subformula over the instance model (L.~3). The substitution hides from SMT those subformulas within the expansion whose evaluation will not be affected by SMT. Subsequently, the algorithm applies expansion and substitution to user-defined operations, taking into account the peculiarities posed by these operations (L.~4-5). Finally, \textbf{deriveFormula} translates the now processed NNF constraint and user-defined operations into an SMT-LIB formula (L.~6).

In the remainder of this section, we elaborate the four algorithms used by Alg.~\ref{algMain} for (1) expansion, (2) substitution, (3) user-operation handling, and (4) SMT-LIB translation, respectively.

\begin{algo}[t]
\caption{Expand Quantifiers Involving Primitive Attributes (\textbf{expand}) \label{algExpand}}
\SetAlgoLined
\setstretch{0.8}
\DontPrintSemicolon 
\SetInd{0.4em}{0.4em}
\SetKwInOut{Parameter}{Fun. calls}
 \SetKwInOut{Input}{Inputs}\SetKwInOut{Output}{Output}
 \setcounter{AlgoLine}{0}
\Input{
(1) An AST node, \var{original}, from the labeled AST of an (NNF) OCL expression;
(2) An instance model, \var{inst}.
}
\BlankLine 
\Output{An AST (root node) resulting from the expansion over \var{inst} of the primitive-attribute-carrying quantifiers of \var{original}.}
\BlankLine 

\Parameter{
\textbf{rename}: replaces within an AST all occurrences of an iterator name by a new name.}

 \BlankLine
 \uIf{(\var{original} is a quantifier node)}
{
Let  \var{expr} be the root AST node for the body of \var{original}\\
 \uIf{(\var{expr} contains some primitive-attribute node)}
{
Let $\mathcal{S}$ be the collection of objects from \var{inst} quantified by \var{original}  \; 
\uIf{($\mathcal{S}$ $\neq$ $\emptyset$)}
{	
$\mathcal{C}\leftarrow\emptyset$\ 
{{\color{darkgreen}\it /* $\mathcal{C}$ will store one copy of \var{expr} per object in $\mathcal{S}$ */}}\\
Let \var{itrName} be the name of the iterator variable of \var{original} \;
\ForEach{($s \in \mathcal{S}$)}
{
Let \var{objId} be the unique identifer for $s$ \;
Let \var{expr\textsubscript{copy}} be a (cloned) copy of \var{expr} \\
$\mathcal{C}$ $\leftarrow$  $\mathcal{C}$ $\cup$ \{\textbf{rename}(\var{expr\textsubscript{copy}}, \var{itrName}, \var{objId})\} 
}
 \uIf{(\var{original} is a \ocl{forAll})}
 {
\var{expansion} $\leftarrow$ $\left(\bigwedge\limits_{j=1}^{j=|\mathcal{C}|} {\mathcal{C}}_{j} \right)$  
  {{\color{darkgreen}\it /* Take the conjunction of the expressions in  $\mathcal{C}$ */}}\\
 }
\Else{ \var{expansion} $\leftarrow$ $\left(\bigvee\limits_{j=1}^{j=|\mathcal{C}|} {\mathcal{C}}_{j} \right)$  
 {{\color{darkgreen}\it /*  Take the disjunction of the expressions in  $\mathcal{C}$ */}}
}
Let \var{root} be the root node of \var{expansion}'s AST \\ 
\Return {\rm \textbf{expand}(\var{root}, \var{inst})}
}
\Else{
 \uIf{(\var{original} is a \ocl{forAll})}
 {
 \Return {\rm an AST node representing \ocl{true}}
 }
\lElse{ \Return {\rm an AST node representing \ocl{false}}}
}
}
\lElse{\Return~ \var{original}}
}
\Else{
\uIf{(\var{original} contains some primitive-attribute node and is further of one of the following types: an equality/inequality operator over collections, \ocl{excludes}, \ocl{excludesAll}, \ocl{includes}, \ocl{includesAll}, \ocl{isUnique}, or \ocl{one})}
{
\nonl\mbox{}{{\color{darkgreen}\it /* \var{original} is an OCL operation with implicit quantification over primitive attributes */}}\\
{\leftskip 0pt\relax Reformulate \var{original} in terms of explicit quantifiers (see Appendix~\ref{sec:expand}), and let \var{root}$'$ be the root node of the AST of the reformulation} \\ 
\Return {\rm \textbf{expand}(\var{root}$'$, \var{inst})}
}
\Else{
\nonl\mbox{}{{\color{darkgreen}\it /* \var{original} involves neither explicit nor implicit quantification. \textbf{expand} is called recursively over \var{original}'s children */}} \\
Let $\mathcal{N}$ be the set of the direct child nodes of \var{original} \; 
\ForEach{($n \in \mathcal{N}$)}
{
$n'$ $\leftarrow$ \textbf{expand}($n$, \var{inst})\\
Replace $n$ with $n'$ in the children of \var{original}\\
}
{\Return {\rm ~ \var{original}}}
}
}
\end{algo}

\begin{figure}[t]
	\centering
	\includegraphics[width=0.75\linewidth]{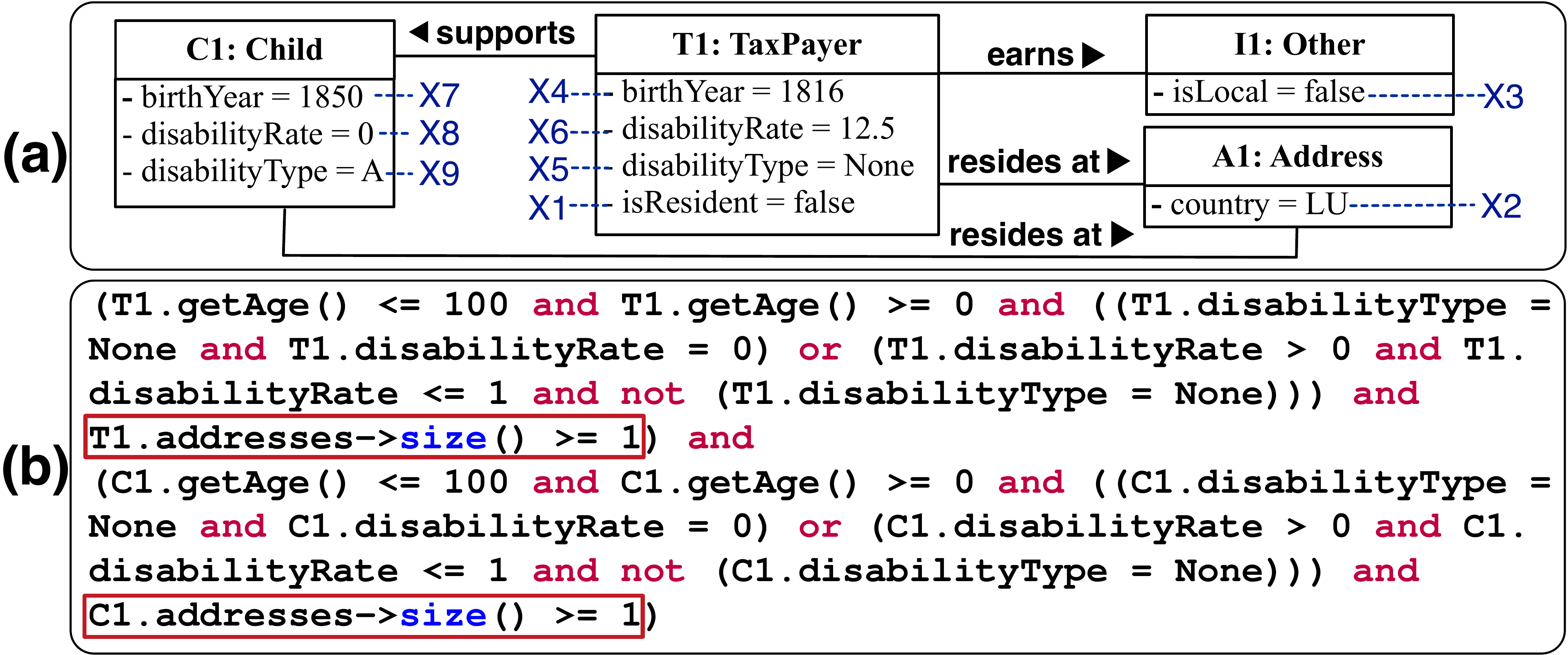}
	\caption{(a) Example Instance Model, and (b) Expansion of S5 of the NNF Constraint in Fig.~\ref{fig:NNF} over the Example Instance Model \label{fig:toSMT-LIB}}
\end{figure}

\sectopic{(1) Expanding the NNF constraint:}
Alg.~\ref{algExpand}, named \textbf{expand}, re-expresses quantifiers in terms of their quantified objects. Expansion is warranted only for nodes whose body contains primitive attributes (L.~1-3). This is because SMT, the way we use it in the PLEDGE approach, cannot affect the evaluation of subformulas that have no primitive attributes. For example, consider S2 and S3 in the NNF constraint of Fig.~\ref{fig:NNF}. The quantifiers in these subformulas do not need to be expanded. A universal quantifier (\ocl{forAll}) is expanded by creating a copy of its body for each quantified object (L.~1-11), and then taking the conjunction of the copies (L.~12-13). As an example, Fig.~\ref{fig:toSMT-LIB}(b) shows S5 of the NNF constraint in Fig.~\ref{fig:NNF} expanded over the instance model of Fig.~\ref{fig:toSMT-LIB}(a). An existential quantifier (\ocl{exists}) is expanded similarly, but with a disjunction applied to the copies (L.~14-15). If the collection of quantified objects is empty, the quantifier in question is replaced with \ocl{true} in case of \ocl{forAll} (L.~19-20), and with \ocl{false} in case of \ocl{exists} (L.~21). 

In addition to the explicit universal and existential quantifiers, expansion applies to certain other (built-in)  OCL operations which, logically speaking, are shortcuts for expressions with quantification (L.~24-26 in Alg.~\ref{algExpand}). To illustrate, the operation \ocl{c->\makeBlue{includes}(e)}, where \ocl{c} is a collection and \ocl{e} is an object, is equivalent to: \ocl{c->\makeBlue{exists}(i\,$\mid$\,i = e)}. We expand such shortcuts by making quantification explicit (L.~25-26 in Alg.~\ref{algExpand}). In Appendix~\ref{sec:expand}, we provide a complete list of these shortcuts alongside their equivalent expressions with explicit quantification. In the final segment of the algorithm (L.~28-31), expansion is applied recursively to the children of a visited node.

\begin{algo}[b]
\caption{Substitute Subformulas Delegated Exclusively to Search (\textbf{substitute})\label{algSubstitute}}
\SetAlgoLined
\setstretch{0.8}
\DontPrintSemicolon 
\SetInd{0.4em}{0.4em}
\SetKwInOut{Parameter}{Fun. calls}
 \SetKwInOut{Input}{Inputs}\SetKwInOut{Output}{Output}
 \setcounter{AlgoLine}{0}
\Input{
(1) An AST node, \var{original}, from the labeled and expanded AST of an OCL expression;
(2) An instance model, \var{inst}.
}
\BlankLine 
\Output{An AST (root node) resulting from the substitution by their concrete evaluation over \var{inst} of all subformulas that need to be hidden from SMT.}

\BlankLine 

\uIf{({\var{original} is labeled \textit{search}})}
{
Let \var{eval} be the result of evaluating over \var{inst} the expression represented by \var{original}\\
\Return the root node of the AST for \var{eval}
}
\Else{
Let $\mathcal{N}$ be the set of the direct child nodes of \var{original} \; 
\ForEach{($n \in \mathcal{N}$)}
{
$n'$ $\leftarrow$ \textbf{substitute}($n$, \var{inst})\\
Replace $n$ with $n'$ in the children of \var{original}\\
}
{\Return {\rm ~ \var{original}}}
}
\end{algo}

\sectopic{(2) Substituting subformulas that are exclusive to search:}
After expansion, Alg.~\ref{algSubstitute}, named \textbf{substitute}, hides from SMT the subformulas that are exclusive to search. Specifically, Alg.~\ref{algSubstitute} substitutes by its concrete evaluation over the instance model any subformula whose root AST node has been labeled \textit{search} by the procedure of Section~\ref{sec:share} (L.~1-3).
For example, consider the already expanded formula in Fig.~\ref{fig:toSMT-LIB}(b). The AST nodes representing the two subformulas delimited by boxes have their root AST node labeled \textit{search} (not shown in the figure). The \textbf{substitute} algorithm replaces these two subformulas with \ocl{true}, noting that \textit{T1} and \textit{C1} in Fig~\ref{fig:toSMT-LIB}(a) are each associated to one instance of \textit{Address}. 
Similarly to the \textbf{expand} algorithm (Alg.~\ref{algExpand}) discussed earlier, the final segment of the \textbf{substitute} algorithm (L. 5-8) recursively applies substitution to the children of the visited node.

Going back to the NNF constraint of Fig.~\ref{fig:NNF}, the expansion and substitution processes would unwind over the NNF constraint of Fig.~\ref{fig:NNF} as follows: S1 to S3 are substituted by their concrete evaluation over the instance model without being expanded. The primitive-attribute-carrying quantifiers in S4 and S5 are first expanded over the instance model of Fig~\ref{fig:toSMT-LIB}(a) in the manner illustrated for S5 in Fig~\ref{fig:toSMT-LIB}(b). Subformulas whose root AST nodes are labeled \textit{search} are then substituted by their concrete evaluation over the same instance model.

\begin{algo}[!t]
\caption{Prepare User-defined Operations for Translation to SMT-LIB (\textbf{processOperations}) \label{algOperations}}
\SetAlgoLined
\DontPrintSemicolon 
\SetInd{0.4em}{0.4em}
\SetKwInOut{Parameter}{Fun. calls}
 \SetKwInOut{Input}{Inputs}\SetKwInOut{Output}{Output}
 \setcounter{AlgoLine}{0}
\Input{
(1) A set $\mathcal{U}$ of the ASTs of the user-defined operations to process; 
(2) The root AST node, \var{root}, of an (NNF) constraint after expansion and substitution; 
(3) An instance model, \var{inst}.
}
\BlankLine 
\Output{A set of ASTs obtained by applying substitution and expansion to the ASTs in $\mathcal{U}$. }
\BlankLine 
\Parameter{
\textbf{expand}: see Alg.~\ref{algExpand}; 
\textbf{substitute}: see Alg.~\ref{algSubstitute}.
\BlankLine 
}

Let $T$ be the AST rooted at \var{root}\\
$\mathcal{U}$\textsubscript{to process} $\leftarrow$ $\emptyset$ {{\color{darkgreen}\it /* Will store the ASTs of those (user-defined) operations that should be translated to SMT-LIB and thus need processing */}} \\
{$\mathcal{U}$\textsubscript{processed} $\leftarrow$ $\emptyset$ {{\color{darkgreen}\it /* Will store the processed ASTs of the operations */}}} \\
 \ForEach{($u$ $\in$ $\mathcal{U}$)}
{
 \uIf{(some node $n$ in $\{T\}$ $\cup$ $\mathcal{U}$ calls the operation for which $u$ is the AST and $n$ is labeled \textit{SMT})}
{
{\leftskip 0pt\relax Let $u$\textsubscript{externalized} be the AST resulting from transforming the primitive attributes  within $u$ into additional parameters for the operation corresponding to $u$}\\
In $\{T\}$ $\cup$ $\mathcal{U}$, replace references to $u$ with references to $u$\textsubscript{externalized}\\
$\mathcal{U}$\textsubscript{to process} $\leftarrow$ $\mathcal{U}\textsubscript{to process} \cup \{u\textsubscript{externalized}\}$
 } 
}
Let $\mathcal{L}$ be an (ordered) list resulting from sorting $\mathcal{U}\textsubscript{to process}$ such that no $u_i\in\mathcal{U}\textsubscript{to process}$ depends on $u_j\in\mathcal{U}\textsubscript{to process}$ where $j>i$

\ForEach{($l$ $\in$ $\mathcal{L}$)}
{

 \uIf{($l$ contains some node that needs to be either expanded or substituted)}
{
Let $\mathcal{S}$ be the set of objects in \var{inst} that call the operation for which $l$ is the AST \\ 
\ForEach{($s$ $\in$ $\mathcal{S}$)}
{
Let \var{op}$_s$ be an $s$-specific (renamed) copy of the operation corresponding to $l$\\
Associate to \var{op}$_s$ as its AST a (cloned) copy of $l$ \\
In $\{T\}$ $\cup$ $\mathcal{L}$, replace references to $l$ with references to \var{op}$_s$\\
\var{op}\textsubscript{expanded} $\leftarrow$ \textbf{expand}(root AST node of \var{op}$_s$, \var{inst})\\
\var{op}\textsubscript{substituted} $\leftarrow$ \textbf{substitute}(\var{op}\textsubscript{expanded}, \var{inst})\\
{$\mathcal{U}$\textsubscript{processed} $\leftarrow$ $\mathcal{U}$\textsubscript{processed}  $\cup$ $\{\text{AST rooted at \var{op}\textsubscript{substituted}}\}$} \\
}
}
\lElse{
{$\mathcal{U}$\textsubscript{processed} $\leftarrow$ $\mathcal{U}$\textsubscript{processed}  $\cup$  $\{l\}$  }
}
}
\Return{\rm $\mathcal{U}$\textsubscript{processed} } \\
\end{algo}

\sectopic{(3) Handling user-defined operations:} User-defined operations may too contain quantification as well as subformulas that should be hidden from SMT. Since user-defined operations are expressed in OCL, we process them in the same way as the NNF constraint, i.e., by invoking \textbf{expand} and \textbf{substitute} (Algs. \ref{algExpand} and \ref{algSubstitute}). Nevertheless, before expansion and substitution can be applied to user-defined operations, some pre-processing is required. Alg.~\ref{algOperations}, named \textbf{processOperations} and discussed next, tailors expansion and substitution to user-defined operations.

Initially, Alg.~\ref{algOperations} determines which user-defined operations require expansion and substitution (L.~1-5). Specifically, these are the operations that are either directly or indirectly used by the (processed) NNF constraint via an AST node labeled \textit{SMT} (L.~4-5). Indirect usage means that some operation \textit{op} is not called directly within the NNF constraint; however, \textit{op} appears in some chain of user-defined operation calls originating from the NNF constraint.

\begin{figure}[t]
	\centering
	\includegraphics[width=0.5\linewidth]{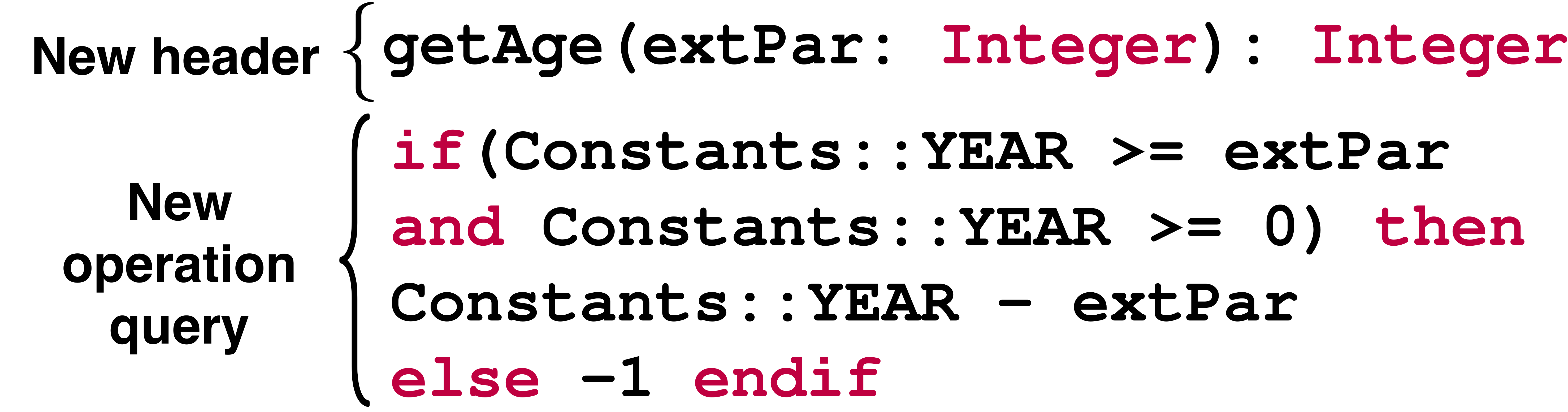}
	\caption{The \textit{getAge} Operation after externalizing Primitive Attributes as Parameters\label{fig:newAge}}
\end{figure}

Due to OCL's object-oriented nature, user-defined operations have access to properties that are encapsulated within the operations' calling objects. For example, the \textit{getAge} operation in Fig.~\ref{fig:example}(a) refers to the \textit{birthYear} attribute of its calling object. In SMT-LIB, however, only the input parameters and the built-in SMT-LIB functions are accessible within the body of a given operation (function in SMT-LIB). To get around this difference between OCL and SMT-LIB, \textbf{processOperations} externalizes  as a parameter any mutable primitive attribute within the user-defined operations that need to be processed (L.~6-8 in Alg.~\ref{algOperations}). For example, the operation call \ocl{T1.getAge()} in the constraint of Fig.~\ref{fig:toSMT-LIB}(b) becomes \ocl{T1.getAge(T1.birthYear)}; the OCL query that defines the operation is also updated accordingly. The \textit{getAge} operation after externalization is provided in Fig.~\ref{fig:newAge}.

The \textbf{expand} and \textbf{substitute} algorithms need to be able to dynamically evaluate OCL subformulas. In the case of \textbf{expand}, this ability is required for extracting the quantified objects, and in the case of \textbf{substitute} -- for calculating concrete results to replace the subformulas that should be hidden from SMT. To enable dynamic (on the fly) evaluations, we need knowledge of the calling object and parameter values of a given operation at the time the operation is processed. This knowledge is available \textit{a priori} for the operations that are called directly by the (processed) NNF constraint. However, when an operation is called only indirectly, the knowledge needs to be obtained from the underlying chain of operation calls. To ensure that we have the required knowledge at the time we need it (during expansion or substitution), the dependencies of each user-defined operation need to be processed before the operation itself. In \textbf{processOperations}, this is achieved by ordering the user-defined operations to process according to their dependencies. Specifically, an operation in position $i$ should not depend on an operation in position $j > i$ (L.~9 in Alg.~\ref{algOperations}). 

To illustrate, let us hypothetically assume that the AST in Fig.~\ref{fig:NNF} is that of a user-defined operation~\textit{op}, instead of that of an NNF constraint. In this case, \textit{op} must be processed first to enable the identification of the objects and parameters of calls to \textit{getAge}. The reverse order, i.e., processing \textit{getAge} before \textit{op}, would not be an option because we neither know the calling object, nor the parameter value to pass to the (externalized version of) \textit{getAge} shown in Fig.~\ref{fig:newAge}.

\begin{algo}[!t]
\caption{Translate to SMT-LIB (\textbf{OCL2SMT-LIB}) \label{algToSMT-LIB}}
\SetAlgoLined
\DontPrintSemicolon 
\SetKwInOut{Parameter}{Fun. calls}
 \SetKwInOut{Input}{Inputs}\SetKwInOut{Output}{Output}
 \setcounter{AlgoLine}{0}
\Input{
(1) The root AST node, \var{root}, of a processed NNF constraint;
(2) A set $\mathcal{U}$ of user-defined-operation ASTs produced by Alg.~\ref{algOperations}.
}
\BlankLine 
\Output{An SMT-LIB translation of the constraint represented by \var{root}.}
\BlankLine 
\Parameter{\textbf{nameOf}: returns the name of a UML element; 
\textbf{SMT-LIBTypeOf}: returns the SMT-LIB counterpart type for a primitive UML type. For example, \textbf{SMT-LIBTypeOf}("\texttt{Boolean}") returns "\texttt{Bool}"; \textbf{sortByDependency}: sorts a set of operations such that the operation in position $i$ does not depend on any operation in position $j>i$; 
\textbf{declareOperation}: defines an OCL operation in SMT-LIB (signature and body). For user-defined operations, this function uses \textbf{toSMT-LIB} (defined below) for generating the SMT-LIB operation body; 
\textbf{toSMT-LIB}: translates OCL to SMT-LIB through the rules in Table~\ref{tab:rules} of Appendix~\ref{sec:toSMT-LIB}. This function further dynamically creates the variable bindings required for lifting any satisfying assignment found by SMT back to the instance model.
}
 
 \BlankLine

 \var{res} $\leftarrow$ "\texttt{(set-option: produce-models true)}" \\ 
 Let $T$ be the AST rooted at \var{root}\\
 Let $\mathcal{V}$ be the set of AST nodes representing the primitive attributes (variables) that appear in $\{T\} \cup \mathcal{U}$ \\
 Let $\mathcal{E}$ be the set of enumerations used as types for the variables in $\mathcal{V}$ \\
\nonl ~ {{\color{darkgreen}\it /* Declare the required enumerations in SMT-LIB */}} \\
\ForEach{($e \in \mathcal{E}$)}
{
 \var{res} $\leftarrow$ \var{res} + "\texttt{({declare-datatypes} ()((}" + \textbf{nameOf}($e$)\\
 Let $\mathcal{L}$ be the set of literals of $e$ \\
 \ForEach{($\ell \in \mathcal{L}$)}
 {
 \var{res} $\leftarrow$ \var{res} + "\texttt{(}" + \textbf{nameOf}($\ell$) + "\texttt{)}"\\
 }
 \var{res} $\leftarrow$ \var{res} + "\texttt{)))}"\\
}
 \nonl ~ {{\color{darkgreen}\it /* Declare the SMT-LIB variables */}} \\
 \ForEach{($v \in \mathcal{V}$)}
 {
 \var{res} $\leftarrow$ \var{res} + "\texttt{(declare-const }" + \textbf{nameOf}($v$) + " " + \textbf{SMT-LIBTypeOf}($v$) + "\texttt{)}" \\
 }
\nonl ~ {{\color{darkgreen}\it /* Define: (1) all user-defined operations, and (2) the OCL built-in operations without counterparts in SMT-LIB */}} \\
 Let $\mathcal{B}$ be the set of OCL built-in operations that (1) are used within $\{T\} \cup \mathcal{U}$, and (2) have no matching function in SMT-LIB (see Appendix~\ref{sec:toSMT-LIB}) \\
 
Let $\mathcal{O}$ be the union of $\mathcal{B}$ and the set of user-defined operations whose ASTs are in $\mathcal{U}$ \\
$\mathcal{O}$\textsubscript{sorted} $\leftarrow$ \textbf{sortByDependency}($\mathcal{O}$) {{\color{darkgreen}\it /* Determine the order in which the operations should be declared */}} \\
 \ForEach{($o$ $\in$ $\mathcal{O}_\text{sorted}$)}
 {
 \var{res} $\leftarrow$ \var{res} + \textbf{declareOperation}($o$, $\mathcal{U}$) ~ {{\color{darkgreen}\it /* $\mathcal{U}$ is needed for producing the SMT-LIB body of user-defined operations */}} \\
 }
\Return {\rm ~\var{res} + "\texttt{(assert }" + \textbf{toSMT-LIB}(\var{root}) + "\texttt{)(check-sat)}"}
\end{algo} 

The final additional consideration regarding user-defined operations has to do with the fact that the same user-defined operation may be called by different objects. This means that (the AST of) a user-defined operation may expand differently for different calling objects. Further and during the substitution process, the concrete evaluation of the same subformula might yield different results for different calling objects. To handle this, \textbf{processOperations} replaces, for each calling object, the original operation with a new (cloned) operation that is specific to the calling object (L.~10-16). This treatment does not apply to the (externalized) \textit{getAge} operation in Fig.~\ref{fig:newAge}, since this operation has no quantification and has no subformulas that are exclusive to search (L.~11). For the sake of argument, let us hypothetically assume that the query defining \textit{getAge} includes some quantification. Given the instance model of Fig.~\ref{fig:toSMT-LIB}(a), the calls to \textit{getAge} in the constraint of Fig.~\ref{fig:toSMT-LIB}(b) would be replaced by \ocl{getAge\_for\_T1} for \textit{T1} and by \ocl{getAge\_for\_C1} for \textit{C1}, with \textit{getAge\_for\_T1} and \textit{getAge\_for\_C1} being clones of the \textit{getAge} operation (Fig.~\ref{fig:newAge}. After this final treatment, the user-defined operations that need processing are expanded and subsituted in exactly the same manner as the NNF constraint (L.~17-19). 

\begin{figure}[t]
	\centering
	\includegraphics[width=0.75\linewidth]{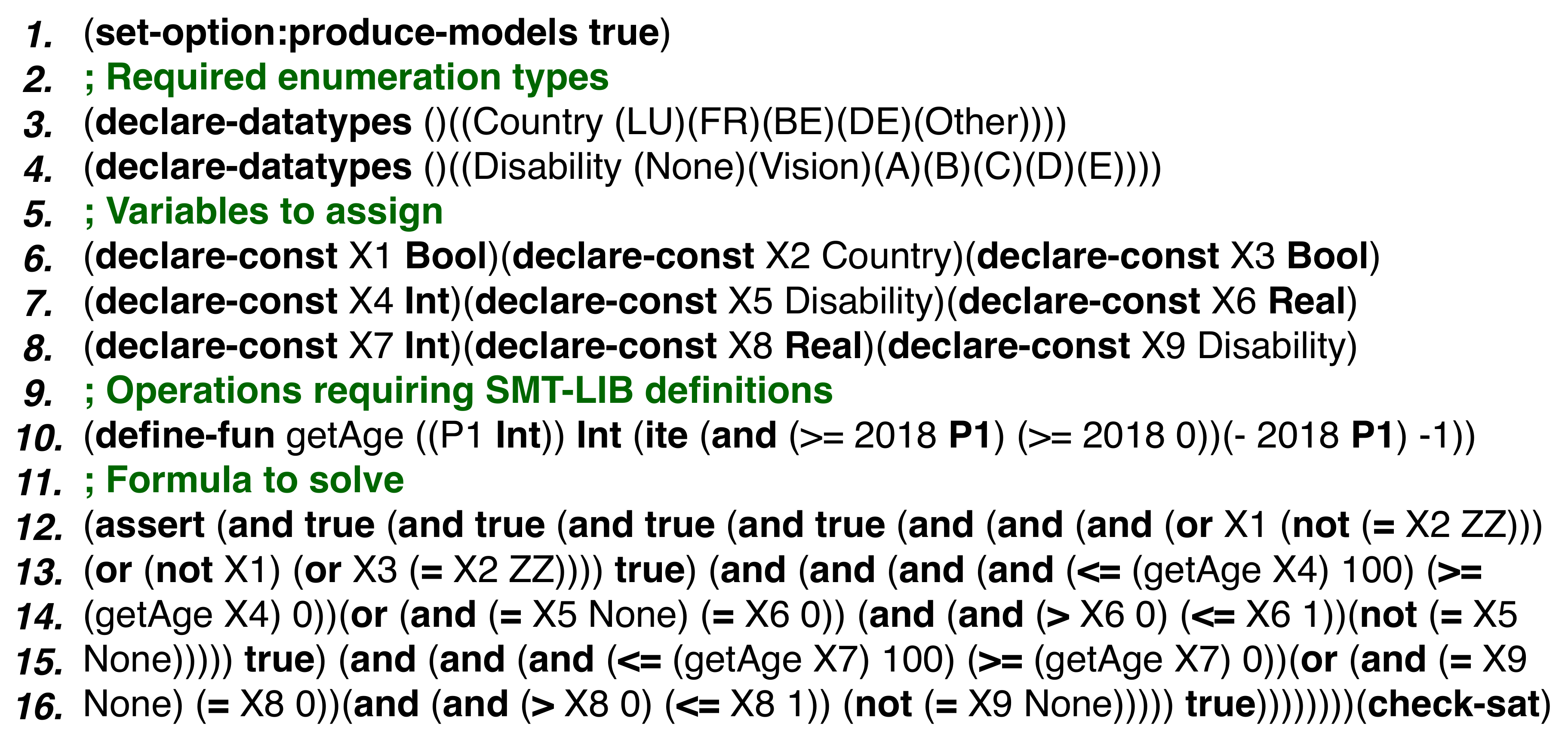}
	\caption{SMT-LIB Formula for the NNF Constraint of Fig.~\ref{fig:NNF} and Instance Model of Fig.~\ref{fig:toSMT-LIB}(a) \label{fig:SMT-LIB}}
\end{figure}

\sectopic{(4) Translation to SMT-LIB:} Alg.~\ref{algToSMT-LIB}, named \textbf{OCL2SMT-LIB}, translates the now processed NNF constraint and user-defined operations to SMT-LIB. Fig.~\ref{fig:SMT-LIB} presents the SMT-LIB formula obtained for the NNF constraint of Fig.~\ref{fig:NNF} and the instance model of Fig.~\ref{fig:toSMT-LIB}(a). L.~1 of the formula of Fig.~\ref{fig:SMT-LIB}, produced by L.~1 of Alg.~\ref{algToSMT-LIB}, indicates to the SMT solver that we would like to obtain a satisfying assignment if the formula is satisfiable. 

L.~2-4 of the SMT-LIB formula in Fig.~\ref{fig:SMT-LIB}, produced by L.~2-10 of Alg.~\ref{algToSMT-LIB}, declare the inv fML enumeration types in SMT-LIB. 
Specifically, L.~2-4 of Alg.~\ref{algToSMT-LIB} extract all UML enumeration \emph{types} that are referenced in the processed NNF constraint and user-defined operations.
Our running example in Fig.~\ref{fig:example} includes two enumerations, namely,  \emph{country} and \textit{Disability}. Since both enumerations are relevant to the processed NNF constraint,  L.~5-10 of Alg.~\ref{algToSMT-LIB} convert them, alongside their enumeration items, to SMT-LIB enumerations as shown in L.~2-4 of the formula.

 L.~5-8 of the formula, produced by L.~11-12 of Alg.~\ref{algToSMT-LIB}, declare the primitive SMT-LIB variables that can be assigned by the SMT solver. 
These variables come from converting to SMT-LIB each primitive attribute of the instance model of Fig.~\ref{fig:toSMT-LIB}(a) that is referenced (directly or indirectly) by the constraint at hand. For example, all attributes in the instance model in  Fig.~\ref{fig:toSMT-LIB}(a) are concerned by the translation since they are of primitive types and are accessible from the processed NNF constraint. The rules for declaring SMT-LIB variables are simple and can be found in Table~\ref{tab:rules} of Appendix~\ref{sec:toSMT-LIB}. For example, and as shown by the SMT-LIB formula in Fig.~\ref{fig:SMT-LIB}, the attributes \emph{isResident} and \emph{country} of the instance model  in  Fig.~\ref{fig:toSMT-LIB}(a)  translate to \ocl{(\makePurple{declare-const} X1 \makePurple{Bool})} and \ocl{(\makePurple{declare-const} X2 Country)}, respectively.  
We note that, as shown by the annotations in Fig.~\ref{fig:toSMT-LIB}(a), each declared SMT-LIB variable is bound to its corresponding primitive attributes in the instance model, e.g., \texttt{X1} and \texttt{X2} respectively refer to the \emph{isResident} and \emph{country} attributes. 
 Later, if the solver finds a solution, these bindings will be used for lifting back satisfying assignments to the instance model.

L.~9-10 of the formula, produced by L.~13-17 of Alg.~\ref{algToSMT-LIB}, define the required built-in and user-defined OCL operations in SMT-LIB. For the NNF constraint in Fig.~\ref{fig:NNF}, only \textit{getAge} needs to be defined. The generation of the SMT-LIB header for \textit{getAge} is straightforward (L.~17 of Alg.~\ref{algToSMT-LIB}). The body of the corresponding SMT-LIB function is built  by conducting a depth-first traversal over the AST of \textit{getAge} (shown in Fig.~\ref{fig:newAge}) and applying a set of predefined OCL to SMT-LIB translation rules (L.~17 of Alg.~\ref{algToSMT-LIB}). For example,  the if-condition \ocl{(Constants::YEAR >= extPar \makePurple{and} Constants::YEAR >= 0)} in  \textit{getAge} of Fig.~\ref{fig:newAge} becomes \ocl{(and (>= 2018 P1) (>= 2018 0))} in the SMT-LIB formula of Fig.~\ref{fig:SMT-LIB}, with \texttt{P1} being a parameter of the SMT-LIB function for \textit{getAge} and \ocl{2018} being the constant value for the static attribute \emph{YEAR} of the class \textit{Constants} in Fig.~\ref{fig:example}. Our complete list of OCL to SMT-LIB translation rules are provided in Appendix~\ref{sec:toSMT-LIB}. 

Several built-in OCL operations such as \ocl{abs}, \ocl{div}, and \ocl{mod} are available as default functions in SMT-LIB. Such operations do not require explicit SMT-LIB function definitions. Built-in OCL operations without a counterpart SMT-LIB function, e.g., \ocl{max} and \ocl{round}, are defined, when needed, using the rules listed in Appendix~\ref{sec:toSMT-LIB} (L.~13-17 of Alg.~\ref{algToSMT-LIB}). For example, \ocl{max} in OCL is translated into the following SMT-LIB function: \ocl{(\makePurple{define-fun} max ((a \makePurple{Real})(b \makePurple{Real})) \makePurple{Real} (\makePurple{ite} (>= a b) a b))}.

L.~11-16 of the formula in Fig.~\ref{fig:SMT-LIB}, produced by L.~18 of Alg.~\ref{algToSMT-LIB}, represent the core assertion derived from the processed NNF constraint. This assertion is produced by applying the OCL to SMT-LIB translation rules listed in Appendix~\ref{sec:toSMT-LIB}. Note that these rules are the same as those used for translating user-defined operation into SMT-LIB.

\subsubsection{Invoking the SMT Solver and Updating the Instance Model} Given the binding established between the variables in the constructed SMT-LIB formula and the primitive attributes in the instance model (see Fig.~\ref{fig:toSMT-LIB}(a)), updating the instance model is straightforward. For example, when invoked over the SMT-LIB formula of Fig.~\ref{fig:SMT-LIB}, the solver may return the following satisfying assignment: {\texttt{X1}\,$\leftarrow$ \texttt{false}}, {\texttt{X2}\,$\leftarrow$ \texttt{FR}}, {\texttt{X3}\,$\leftarrow$ \texttt{false}}, {\texttt{X4}\,$\leftarrow$ \texttt{1918}}, {\texttt{X5}\,$\leftarrow$ \texttt{A}}, {{\texttt{X6}\,$\leftarrow$ \texttt{1}}}, {\texttt{X7}\,$\leftarrow$ \texttt{1975}}, {\texttt{X8}\,$\leftarrow$ \texttt{1}}, and {\texttt{X9}\,$\leftarrow$ \texttt{A}}. Using the binding shown in Fig.~\ref{fig:toSMT-LIB}(a), the instance model is updated by lifting back this satisfying assignment. In our example, the updated instance model is already a valid solution and thus returned to the user.

\section{Evaluation}\label{sec:evaluation}
In this section, we evaluate the PLEDGE  approach over three industrial case studies. 
\subsection{{Research Questions (RQs)}}\label{sec:RQs}

\sectopic{RQ1. How does PLEDGE  fare against the state of the art in terms of success rate and execution time? }
We examine in RQ1 whether PLEDGE  presents practical improvements over Alloy~\cite{AlloyTool}, UMLtoCSP~\cite{UMLtoCSPTool}, and pure search as implemented by the baseline we build on~\cite{SoltanaTool} (see Section~\ref{sec:search}).

\sectopic{RQ2. Can PLEDGE generate large instance models in practical time?}
The need to test a large variety of system scenarios and system robustness and performance, e.g., load and stress testing~\cite{Zhang:11}, entails large test inputs. RQ2 investigates how PLEDGE  scales as we have it produce increasingly larger instance models.

\sectopic{RQ3. Does PLEDGE offer any scalability advantage over the alternatives, when PLEDGE is not the obvious best choice according to the results of RQ1?}
The goal of this RQ, posed as a follow-on to RQ1, is to provide insights about the scalability of PLEDGE in comparison to Alloy, UMLtoCSP, and the baseline. As we explain in our answer to RQ1, these alternatives are equally good or better than PLEDGE in one out of our three case studies. For that particular case study, RQ3 contrasts PLEDGE's scalability trends seen in RQ2 against those of the alternatives. Intuitively, RQ3 investigates whether PLEDGE can provide an advantage over the alternatives for building \emph{large} instance models in the case study where PLEDGE has no benefit over the alternatives otherwise.
 
\subsection{Tool Support}\label{sec:implementation}
\begin{figure}[b]
	\centering
	\includegraphics[width=0.8\linewidth]{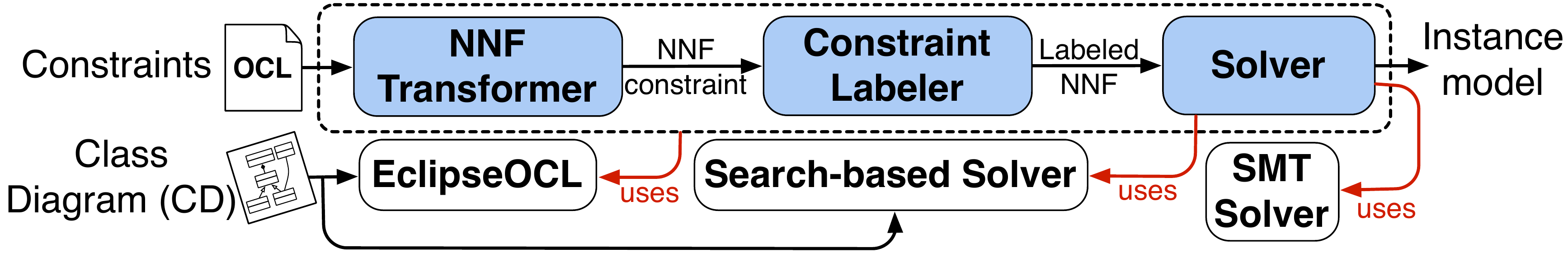}
	\caption{PLEDGE's Architecture (new modules implemented as part of this article are shaded blue)\label{fig:tool}}
\end{figure}

Fig.~\ref{fig:tool} provides an overview of the PLEDGE tool. \emph{NNF Transformer} turns the input OCL constraints into an NNF constraint (Section~\ref{sec:ocl}). \emph{Constraint Labeler} spreads the solving tasks over search and SMT (Section~\ref{sec:share}). 
\emph{Solver} implements the three iterative steps depicted in Fig.~\ref{fig:process} (Sections~\ref{sec:search} through \ref{sec:z3}). 
The PLEDGE tool uses Eclipse~OCL~\cite{EclipseOCL} for building ASTs and evaluating OCL expressions, the OCL solver from our previous work~\cite{Soltana:2017} for search-based solving, and Z3~\cite{Z3Tool} for SMT solving.
Excluding comments and third-party libraries,  the PLEDGE tool consists of $\approx$16K lines of Java code.
The tool is available online at \href{https://sites.google.com/view/hybridoclsolver/}{https://sites.google.com/view/hybridoclsolver/}. 

\begin{table}[b]
\vspace{1em}
\caption{Statistics for the Case Studies}\label{tab:stat}
\vspace{-0.5em}
\fontsize{7}{12}\selectfont
\begin{tabular}{|p{10pt}p{5pt}|p{5.5cm}|S[
		table-number-alignment = right,
		table-column-width = 0.1pt,
		group-digits=true]|
S[
		table-number-alignment = right,
		table-column-width = 0.1pt]
|S[
		table-number-alignment = right,
		table-column-width = 0.1pt]
	|}
\cline{4-6}

\multicolumn{1}{p{ 10pt}}{} & \multicolumn{1}{p{ 5pt}}{} & \multicolumn{1}{p{ 5.5cm}|}{} & \textbf{Case A}& \textbf{Case B} & \textbf{Case C} \tabularnewline
\cline{1-6}
 \multicolumn{1}{|p{5pt}}{\multirow{6}{*}{{\rotatebox{90}{{\textbf{Class Diagram~~~~~~~~~~~~~~~~~~~~~~~~~}
}}}}} & \multicolumn{1}{|p{ 5pt}|}{ 1} & \multicolumn{1}{p{ 5.5cm}|}{\textrm{\#\,of classes}} &  64 & 57 & 55 \tabularnewline
\cline{2-6}
& \multicolumn{1}{|p{ 5pt}|}{ 2} & \multicolumn{1}{p{ 5.5cm}|}{\textrm{\#\,of generalizations}} & 43& 35 & 20\tabularnewline
\cline{2-6}
& \multicolumn{1}{|p{ 5pt}|}{ 3} & \multicolumn{1}{p{ 5.5cm}|}{\textrm{\#\,of associations }} & 53 & 44& 30 \tabularnewline
\cline{2-6}
& \multicolumn{1}{|p{ 5pt}|}{ \multirow{1}{5pt}{4}} & \multicolumn{1}{p{ 5.5cm}|}{\textrm{\#\,of enumerations (Avg \#~of  literals per enumeration)}} & \multicolumn{1}{p{0.9cm}|}{\multirow{1}{*}{\hbox{17\,(6.11)}\phantom{0}}} &
\multicolumn{1}{p{0.9cm}|}{\multirow{1}{*}{\hbox{\phantom{0}4\,(3.25)}\phantom{0}}}
&
\multicolumn{1}{p{0.9cm}|}{\multirow{1}{*}{\hbox{\phantom{0}8\,(4.87)}\phantom{0}}}
\tabularnewline
\cline{2-6}

& \multicolumn{1}{|p{ 5pt}|}{ \multirow{2}{*}{5}} & \multicolumn{1}{p{ 5.5cm}|}{\textrm{\#\,of attributes  \newline (primitive\,+\,non-primitive) }} & \multicolumn{1}{p{0.9cm}|}{\multirow{2}{*}{
\shortstack[1]{
\phantom{0}\phantom{.}\phantom{0}208\phantom{.}
\\
\hspace{-0.5em}\hbox{(158+50)}\phantom{.}
}
}
}
 &
 \multicolumn{1}{p{0.9cm}|}{\multirow{2}{*}{
\shortstack[1]{
\phantom{0}\phantom{.}\phantom{0}\phantom{.}116
\\
(72+44)
}
}
}
&
 \multicolumn{1}{p{0.9cm}|}{\multirow{2}{*}{
\shortstack[1]{
\phantom{0}\phantom{.}\phantom{0}\phantom{0}255\phantom{.}
\\
\hspace{-0.5em}\hbox{(219+36)}
}
}
} \tabularnewline
\hline
\hline
 \multicolumn{1}{|p{3pt}}{\multirow{5}{*}{{\rotatebox{90}{ {\textbf{NNF Constraint}\,
}}}}} & \multicolumn{1}{|p{ 5pt}|}{ \multirow{1}{5pt}{6}} & \multicolumn{1}{p{ 5.5cm}|}{\textrm{\#\,of nodes in the AST of the  NNF constraint}} &  \multicolumn{1}{p{0.9cm}|}{\multirow{1}{*}{\phantom{0}\phantom{0}\phantom{.}2004}} & 
 \multicolumn{1}{p{0.9cm}|}{\multirow{1}{*}{\phantom{0}\phantom{0}\phantom{.}1844}} 
 & 
  \multicolumn{1}{p{0.9cm}|}{\multirow{1}{*}{\phantom{0}\phantom{0}\phantom{.}4014}} 
  \tabularnewline
\cline{2-6}
& \multicolumn{1}{|p{ 5pt}|}{ 7} &\multicolumn{1}{p{ 5.5cm}|}{\textrm{\#\,of AST nodes labeled \textit{search}}} & 574 & 1128 & 1814 \tabularnewline
\cline{2-6}
& \multicolumn{1}{|p{ 5pt}|}{ 8} &\multicolumn{1}{p{ 5.5cm}|}{\textrm{\#\,of AST nodes labeled \textit{SMT}}} & 1239 & 602 & 2017 \tabularnewline
\cline{2-6}
& \multicolumn{1}{|p{ 5pt}|}{\multirow{1}{ 5pt}{9}}&\multicolumn{1}{p{ 5.5cm}|}{\textrm{\#\,of AST nodes labeled \textit{both}}}& 63 & 9 & 144 \tabularnewline
\cline{2-6}
& \multicolumn{1}{|p{5pt}|}{\multirow{1}{*}{\hspace{-0.2em}10}}&\multicolumn{1}{p{ 5.5cm}|}{\textrm{\#\,of unlabeled AST nodes  (representing constants)}}&  \multicolumn{1}{p{0.9cm}|}{\multirow{1}{*}{\phantom{0}\phantom{0}\phantom{0}\phantom{.}128}}   &  \multicolumn{1}{p{0.9cm}|}{\multirow{1}{*}{\phantom{0}\phantom{0}\phantom{0}\phantom{0}\phantom{.}55}}  &  \multicolumn{1}{p{0.9cm}|}{\multirow{1}{*}{\phantom{0}\phantom{0}\phantom{0}\phantom{0}\phantom{.}39}}  \tabularnewline
\hline
\end{tabular}
\end{table}

\subsection{Description of Case Studies}\label{sec:description}
Our evaluation is based on three industrial case studies from three distinct domains. The case studies are denoted Case~A, Case~B, and Case~C.
The source material for each case study is a test data model expressed using UML using Class Diagrams (CDs) and OCL. Our case study material is borrowed from previous work. 
Specifically, Case~A was built for system testing of an eGovernment application concerned with calculating personal income taxes~\cite{Soltana:2017}, Case~B for system testing of an occupant detection system in cars~\cite{Wang}, and Case~C for system testing of a satellite communication system~\cite{Dan}. Our case study material (in a sanitized form) is available from PLEDGE's website. The example described in Section~\ref{sec:example} and used throughout the article is a simplified excerpt from Case~A.

The existing work strands from which our case study material originates employ different test strategies. In particular, Case~A uses statistical testing~\cite{Runeson:95}, Case-B uses coverage-driven model-based testing~\cite{AliBHP10}, and Case-C uses data mutation testing~\cite{Wong}. Despite the different test strategies, the case studies all share a common technical challenge, which is to find a practical way of producing system test data under OCL constraints. Our evaluation in this article is not meant at assessing the effectiveness of the test strategies in the prior work where our case study material comes from. Instead, we evaluate the applicability and scalability of the PLEDGE  approach for system test data generation, irrespectively of the test strategy.

Table~\ref{tab:stat} provides overall statistics about the CDs and OCL constraints in our case studies. 
Specifically, rows 1 to 5 in Table~\ref{tab:stat} provide various size measurements for the CDs. The three CDs have large and
comparable sizes, with Case A including the most structural features such as associations and generalizations,
and Case C including the highest number of attributes. 
Nevertheless, we note that the constraint statistics (rows 6 to 10 of Table~\ref{tab:stat}) are for the single NNF constraint derived in each case study (see Section~\ref{sec:ocl}), rather than the original constraints. This provides a more convenient basis for comparing the complexity of the solving tasks across the case studies when we discuss the evaluation results next.

 Row~6 in Table~\ref{tab:stat}  shows the number of nodes within the ASTs of the NNF constraints built for Case~A, Case~B, and Case~C. 
To help the reader relate to these numbers, we note that the AST of the NNF constraint in Fig.~\ref{fig:NNF} includes only 253 nodes. 
This means that the NNF constraint of Case~C is $\approx$15 times larger than that of the running example.  
Rows~7 to 10  in Table~\ref{tab:stat}   intuitively provide information about how search and SMT share the solving tasks for Case~A, Case~B, and Case~C.
From these numbers, we observe that the task-split balance between search and SMT differs across the case studies.
Specifically, Case~A and Case~C have more subformulas delegated to SMT than search, with the situation being the opposite for Case~B.

\subsection{Results and Discussion}\label{sec:results}
In this section, we answer the RQs of Section~\ref{sec:RQs} via experimentation. Our experiments were performed on a computer with a 3GHz dual-core processor and 16GB of memory. 
The maximum number of iterations for metaheuristic search was set to 1000. The maximum number of iterations for metaheuristic search was set to 1000. This choice was made based on experimentation with the (purely search-based) baseline solver. In particular, we observed that, over our case studies, the fitness function plateaued within 1000 iterations; in other words, improvements in the fitness function value were insignificant past 1000 iterations. Further, and as importantly, all successful runs of the baseline solver required less than 1000 search iterations. 

\sectopic{RQ1.}
We ran PLEDGE, the baseline search-based solver, UMLtoCSP (Version from 2009 with its default ECLiPSe solver V5.0~\cite{ECLiPSe}), and Alloy (V5.0) on the case studies to which they are applicable. Both Alloy and UMLtoCSP have user-specified parameters for bounding the search space. In Alloy, such bound is provided through the ``scope'' parameter~\cite{JacksonBook2012}. As an example, if the scope is set to five, the Alloy Analyzer will attempt to instantiate each entity at most five times when searching for a solution. UMLtoCSP works in a similar manner, but requires bounds to be specified for individual classes and associations. These bounds have a significant impact on the performance of Alloy and UMLtoCSP. The general recommendation is to use the smallest bounds within which a solution exists~\cite{JacksonBook2012}. In RQ1, and specifically for Case~B where Alloy and UML2CSP are applicable, we set Alloy's scope parameter to two; no solution exists when the scope is set to one. Similarly, for UMLtoCSP, we set to two the instantiation bounds for all classes and associations.
Further, Alloy uses SAT solving as backend, and comes prepackaged with several alternative SAT solvers. We considered two alternatives: SAT4J~\cite{SAT4J} -- Alloy's default SAT solver -- and miniSat~\cite{miniSat} -- the SAT solver recommended in Alloy's documentation for better performance~\cite{AlloyTool}. 
The results we report  in RQ1 and RQ3 on Alloy are the ones obtained using miniSat, noting that Alloy used with miniSat consistently outperformed Alloy used with SAT4J in all our experiments.

PLEDGE and the baseline solver are applicable to all three case studies. Alloy and UMLtoCSP are applicable to Case B and only for a scope higher or equal to two. In RQ1, as mentioned earlier, the scope was set to a minimum of two. Case~A and Case~C have features which are common in practice, but which Alloy and UMLtoCSP do not support. In the case of Alloy, the main issues for which we could not find a workaround had to do with the absence of real / floating-point arithmetic and string operations. In the case of UMLtoCSP, the  main limitation was the lack of support for \ocl{null} and \ocl{OclInvalid}. This meant that several OCL operations used in our case studies, e.g., \ocl{oclIsUndefined} (illustrated in Fig.~\ref{fig:example}(b)) and \ocl{oclIsInvalid}, could not be handled by UMLtoCSP. We note that for Case~B, UML2Alloy~\cite{UML2AlloyTool} (V0.52) turned out to be adequate for translating that case study to Alloy's input language. The results we report here (and for RQ3) for Alloy are based directly on the Alloy model derived by UML2Alloy. The derived model can be found on our tool's website. 

For each solver considered, we measured the execution time. To account for random variation, we ran the solvers 100 times each. We computed the success rate, i.e., how often each solver succeeded in finding a solution, based on the 100 runs.
The CDs in our case studies happen to have a root class, i.e., a class from whose instances one can reach all other objects in an instance model. In Case~A, this class is \textit{Household} --  a unit of taxation (not shown in our example of Fig.~\ref{fig:example}). In Case~B, the class is \textit{BodySenseSystem} -- a container for all system components. In Case C, the class is \textit{Vcdu} -- a container for data packages transmitted over a satellite channel.
Through the notion of non-emptiness constraint discussed in Section~\ref{sec:ocl}, we instructed each applicable solver to generate exactly one instance of the root class. This is a natural choice for comparison purposes, since a single instance of the root class represents the smallest meaningful (system) test input in each of our case studies. Had the case studies not had a root class, we would have chosen multiple classes to instantiate in a way that would match system testing needs. We further note that one can seed PLEDGE and the baseline solver with a partial (and potentially invalid) instance model as the starting point. We did not use this feature; both solvers used the default option, which is to start from an \emph{empty} instance model.

\begin{table}[!t]
\caption{{RQ1 Results}}\label{tab:RQ1}
\centering
\fontsize{7}{12}\selectfont
\setlength{\tabcolsep}{2pt}
\begin{tabular}{p{0.1cm}|p{1.8cm}|p{1.3cm}|p{1.3cm}||p{1.3cm}|p{1.3cm}|p{1.2cm}|p{1.2cm}||p{1.3cm}|p{1.3cm}|}
\cline{3-10}
  \multicolumn{1}{c}{}
  &
 &  \multicolumn{2}{p{2.6cm}||}{ \centering\textbf{Case~A}} &  \multicolumn{4}{p{5cm}||}{ \centering\textbf{Case~B}}&  \multicolumn{2}{p{2.6cm}|}{ \centering\textbf{Case~C}}\tabularnewline
\cline{3-10}

  \multicolumn{1}{c}{}& &
\centering  \textbf{Baseline} &  \centering \textbf{PLEDGE} &  \centering  \textbf{Alloy}  &  \centering  \textbf{UMLtoCSP} & \centering  \textbf{Baseline} & \centering \textbf{PLEDGE} & \centering \textbf{Baseline} & \centering \textbf{PLEDGE}  \tabularnewline
\hline
 \multicolumn{1}{|p{0.1cm}|}{\multirow{2}{0.15cm}{\hspace*{-.0em}1}} &\multirow{2}{*}{ \textrm{Succeeded?}}   &
  \multirow{2}{1.3cm}{28\,/\,100 \newline runs} & 
   \multirow{2}{1.3cm}{100\,/\,100 \newline runs}  &  
   \multirow{2}{1.5cm}{{\color {black}100\,/\,100 \newline runs}}  & 
  \multirow{2}{1.3cm}{28\,/\,100 \newline runs} & 
    \multirow{2}{2cm}{100\,/\,100  \newline runs}  & 
      \multirow{2}{1.7cm}{100\,/\,100 \newline runs} &
  \multirow{2}{1.3cm}{0\,/\,100 \newline runs} 
   &   \multirow{2}{1.3cm}{100\,/\,100 \newline runs}
    \tabularnewline
    
\multicolumn{1}{|p{0.1cm}|} {}& & & &  & &   & &   &       \tabularnewline
\hline

 \multicolumn{1}{|p{0.1cm}|}{\multirow{3}{0.15cm}{\hspace*{-.0em}2}} & \multirow{3}{2.2cm}{\textrm{\color{black}Average \linebreak execution time  }}  & 
  \multirow{3}{2cm}{{587.2 sec} \\
{SD = 383.2}}

 &  

 \multirow{3}{1.62cm}{{27.8 sec} \\
{SD = 18.25}}
 &

       \multirow{3}{2cm}{{{\color{black}0.045 sec} \\
{\color{black}SD = 0.005}}
     }
     
 &  
        \multirow{3}{2cm}{{{\color{black}1.56 sec} \\
{\color{black}SD = 0.055}}
     }
     
 &  
    \multirow{3}{2cm}{{4.5 sec} \\
{SD = 1.7}}
 &  
 
    \multirow{3}{1.5cm}{{3.6 sec} \\
{SD = 0.7}}

 &
   \multirow{3}{2cm}{{1771 sec} \\
{SD = 538.9}}
  & 
     \multirow{3}{2cm}{{44.3 sec} \\
{SD = 7.2}}
  
  \tabularnewline
\multicolumn{1}{|p{0.1cm}|} {}& & & &  & &   & &   &       \tabularnewline
\hline

\multicolumn{1}{|p{0.1cm}|}{\multirow{3}{0.15cm}{\hspace*{-.0em}3}} &  \textrm{\color{black}Average instance model size  (for successful runs)}  &  
 \multirow{3}{1.62cm}{{  28.8 obj.} \\
{SD = 5.8}}
& 
 \multirow{3}{1.62cm}{{  36.6 obj.} \\
{SD = 7.1}}

 & 
 
 \multirow{3}{1.62cm}{{\color{black}  21 obj.} \\
{\color{black}SD = 0}}

 &  
 
 \multirow{3}{1.62cm}{{\color{black}  19 obj.} \\
{\color{black}SD = 0}}

 &  
 
 \multirow{3}{1.62cm}{{  19.7 obj.} \\
{SD = 2.1}}

 &    
   \multirow{3}{1.62cm}{{  23.3 obj.} \\
{SD = 3.1}}
& 
  \multirow{3}{2cm}{--  }

& 
\multirow{3}{2cm}{{  121.4 obj.} \\
{SD = 6.1}}
\tabularnewline
\hline
\end{tabular}
\vspace*{1em}
\end{table}

Table~\ref{tab:RQ1} shows the results of RQ1, alongside the size of the generated instance models. 
Since each solver was executed 100 times, Table~\ref{tab:RQ1} reports averages (Avg) and standard deviations (SD) for the execution time and instance model sizes.

As seen from the table, PLEDGE was applicable to all three case studies and able to produce results in all its runs. 
In Case~B, all four solvers were equally successful and very quick, with Alloy being the quickest. 
How the execution time trends observed over Case~B evolve as one attempts to build larger instance models is assessed in RQ3.

With regard to the performance of PLEDGE compared to the baseline, the results in Table~\ref{tab:RQ1} suggest that PLEDGE brings no practical benefit in Case~B. 
This can be explained as follows: The primitive attributes in Case~B are less constrained than those in Case~A and Case~C (rows~8~and~9 of Table~\ref{tab:stat}). We further observed that most operations on the primitive attributes in Case~B are simple and of the form \ocl{x\,$\langle$op$\rangle$\,constant}, where \ocl{x} is a primitive attribute and \ocl{$\langle$op$\rangle$} is an equality or inequality operator. 
Consequently, in Case~B, search was as effective as SMT in handling primitive attributes.

In Case A and Case C, PLEDGE has better effectiveness (success rate) and efficiency (execution time) compared to the baseline. As shown by the first row of Table~\ref{tab:RQ1}, PLEDGE was able to produce valid instance models in all its runs. In contrast, the baseline solver produced valid instance models in only 28\% of the runs in Case A and in none of the runs in Case C. Specifically, in Case~A, PLEDGE outperformed the baseline by a factor of $\approx$3.5 in terms of success rate and a factor of $\approx$21.7 in terms of  execution time. In Case~C, only PLEDGE was able to generate data. Further, and as indicated by the second row of Table~\ref{tab:RQ1}, PLEDGE has a smaller execution time than the baseline for Case~A and Case~C: in Case A, the average execution time of PLEDGE was 27.8 seconds versus 587.2 seconds for the baseline, and, in Case C -- 44.3 seconds versus 1771 seconds.

Statistically speaking, in terms of success rate, PLEDGE has significant benefits over the alternatives for Case~A and Case~C. In particular, and according to the statistical Z-test for the significance of proportions, p-values are <$10^{-5}$. 
Recall from Table~\ref{tab:RQ1} that for Case~B, all the tools have a 100\% success rate. 
As for the execution time, PLEDGE is significantly better than the baseline in Case A and Case C, recalling that Alloy and UMLtoCSP are not applicable to these two case studies. In particular, the p-values from a t-test over the execution time distributions of PLEDGE and the baseline are <$2.2^{-16}$  and <$3.268^{-15}$ for Case~A and Case~C, respectively. In Case~B, Alloy has a significantly better execution time; p-values are <$2.2^{-16}$ against PLEDGE, \hbox{UMLtoCSP, and the baseline.}

Although Case~A and Case~B are similar with respect to the size of the NNF constraint (row 6 of Table~\ref{tab:stat}) and the size of the generated data (row 3 of Table~\ref{tab:RQ1}), the performance difference between PLEDGE and the baseline solver is much larger in Case~A than in Case~B. The better performance of PLEDGE in Case~A is due to SMT being more efficient than search at handling linear arithmetic, which is not only more prevalent in Case~A, but also more complex. Notably, Case~A has numerous inequalities with more than one primitive attribute and \hbox{using OCL operations such as \ocl{max} and \ocl{round}.}

Case~C is inherently more complex than Case~A and Case~B, as suggested by Tables~\ref{tab:stat} and \ref{tab:RQ1}. 
The additional complexity has a detrimental effect on the baseline solver, resulting in all runs of the baseline solver on Case~C to be unsuccessful. This outcome does \emph{not} imply that the baseline solver is theoretically bound to fail on Case~C. Rather, the outcome is a strong indication of the baseline solver being inefficient for Case~C.

\vspace*{.5em}
\begin{mdframed}[style=MyFrame]
\it
The answer to \textbf{RQ1} is that, for one of our case studies, Case~B, which has comparatively simple constraints, all the techniques considered were as effective, and produced results within seconds. For the other two case studies, Case~A and Case~C, Alloy and UMLtoCSP were not applicable, thus reducing the set of alternatives to pure search and our hybrid approach.
How PLEDGE fares against pure search depends on the complexity of the constraints over primitive attributes. 
If such constraints are few and simple, both approaches have similar performance. 
But, when primitive attributes are subject to many complex restrictions, the PLEDGE approach performs substantially better.
In Case~A, PLEDGE was $\approx$3.5 times more successful and $\approx$21.7 faster than \hbox{pure search. In Case~C, only PLEDGE yielded results.}
\end{mdframed}

\sectopic{RQ2.} 
We measured the execution time of PLEDGE for building progressively larger instance models -- a necessity for system testing, as noted earlier. Whereas in RQ1 we set to one the number of instances of the root class, in RQ2, we initially set this number to 20 and increased it up to 100 in increments of 20. Due to the nature of the root class (see RQ1), increasing its number of instances has a knock-on effect, leading to an across-the-board increase in the size of the instance model. 

This knock-on effect comes primarily from the fact that most NNF-constraint fragments are of the shape \ocl{Context.\makeBlue{allInstances}()->\makeBlue{forAll}(condition)}, potentially with several nested layers of quantification. The
more instances of \ocl{Context} there are, the more {computationally intensive} solving becomes, particularly the expansion and fitness function calculations. 

Despite the generated data having multiple roots, it is important to note that the data is not necessarily the union of independently constructed single-rooted instance models. This is because the roots are often logically connected. For a simple example from Case~A, taxpayers are required to have distinct identifiers; this induces a global relationship between all taxpayers in all households (roots). For another example from the same case, consider child support which plays a role in several of the constraints. Case~A allows children to be supported by taxpayers from different households; this is necessary, for example, when the parents of a child are divorced, with the child being in one household and the (financially) supporting parent in another. To build a realistic instance model, we need to cover such situations and thus be able to create links between objects falling under different roots. {In short, building an instance model with multiple roots is not always decomposable into simpler solving tasks.}

As a final remark, we note that, although this has no bearing on RQ2, we did not attempt to maximize diversity when creating multiple class instances in the instance model. If necessitated by the test strategy, diversity can be effectively enforced by dynamically updating the OCL constraints with new inequalities to guide data generation toward diversity~\cite{Soltana:2017}. 

\begin{figure}[b]
	\centering
	\includegraphics[width=0.87\linewidth]{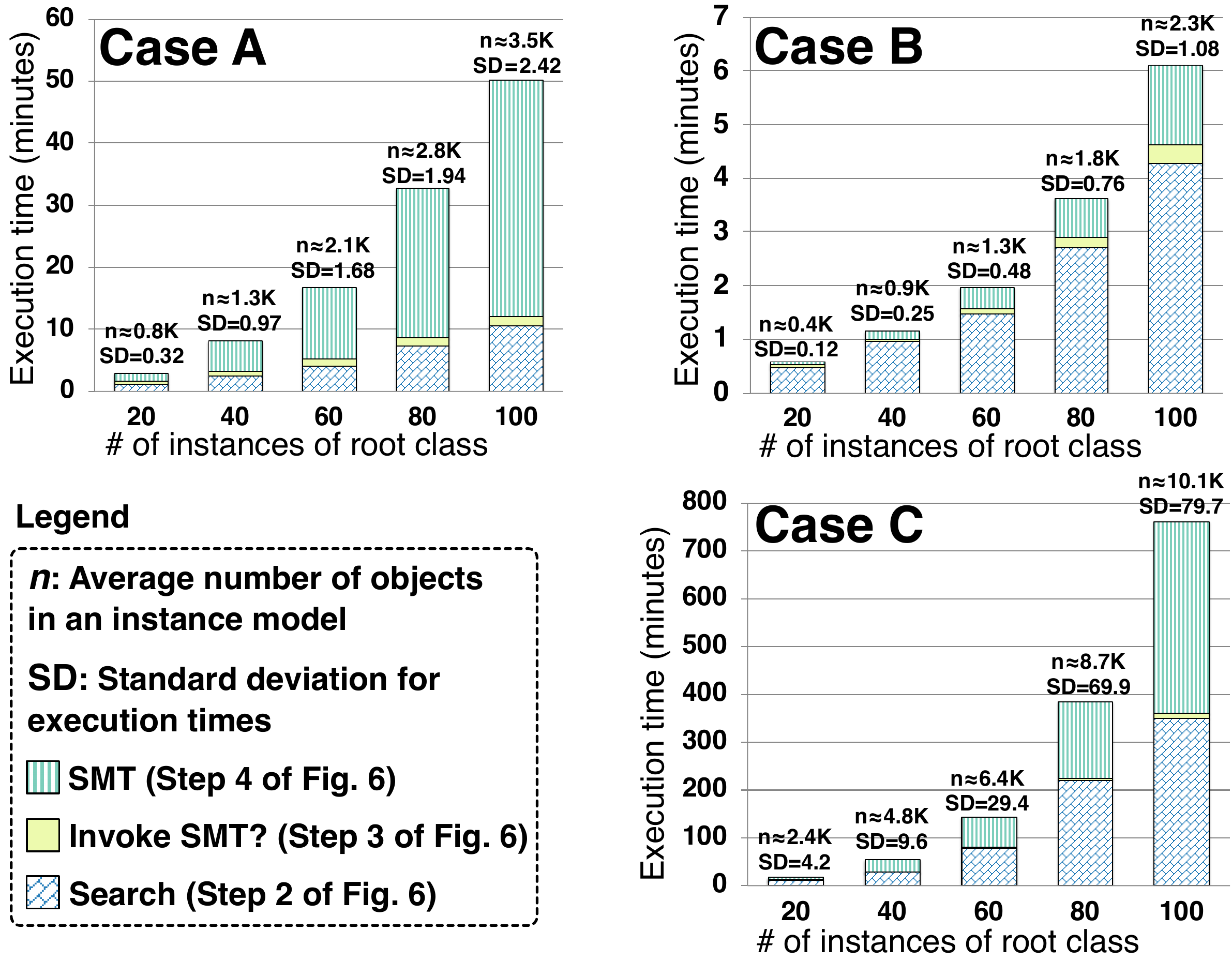}
	\caption{Execution Times for PLEDGE (RQ2)\label{fig:RQ2}}
	\vspace*{1em}
\end{figure}

Fig.~\ref{fig:RQ2} shows for each case study the execution time of PLEDGE over instance models containing 20 to 100 instances of the root class.
To account for random variation, we calculated average execution times based on ten runs. For example, and as shown in Fig.~\ref{fig:RQ2}, the average execution time for generating an instance model with 60 instances of the root class  for Case~A, Case~B, and Case~C is 16.79 minutes (SD=29.4), 1.79 minutes (SD=0.84), and 143.64 minutes (SD=79.7), respectively. The average execution time for Case~C is the highest, while that of Case~B is the lowest. This is because, as we explain in RQ1,  the constraints in Case~C are inherently more complex than those in Case~A and Case~B (see rows 6 to 10 in Table~\ref{tab:stat}), while those in Case~B are the simplest, i.e., mostly simple inequalities over Boolean attributes. 

In Fig.~\ref{fig:RQ2}, we further report the average number of objects in the generated instance models ($n$).
For example, on average, the largest instance models generated for Case~A, Case~B, and Case~C contain 3.5K, 2.3K, and 10.1K~objects, respectively. Although  Case~C has fewer classes and associations than Case~A and Case~B (rows 1 and 3 of Table~\ref{tab:stat}), it requires the generation of a significantly higher number of objects to satisfy the multiplicity and user constraints (encoded in the NNF constraint). Subsequently, and as shown in Fig.~\ref{fig:RQ2}, the execution times measured for Case~C are significantly higher than those of Case~A and Case~B. In general, to various degrees but across all studies, the increase in execution times is exponential.

In addition, Fig.~\ref{fig:RQ2} provides a breakdown of the execution times over the iterative steps of our hybrid solving process, i.e., Steps 2 through 4 of Fig.~\ref{fig:process}. For example, generating an instance model with 60 instances of the root class for Case C took in total 143.63 minutes, with 76.58, 2.11, 64.94 minutes spent on Steps 2, 3, and 4, respectively. 
 The breakdown indicates that the time required for Step 3 (checking whether SMT should be invoked) is negligible. The proportions of the execution times for Steps 2 and 4 reflect how the solving load is spread over search and SMT, taking into account both the characteristics of the NNF constraint to solve and the size of the instance model. Across all the $3\times5\times10=150$ runs underlying the results of Fig.~\ref{fig:RQ2}, 79.5\% of the execution time in Step~4 was spent on expansion and substitution, 0.7\% on SMT solving by Z3, and the remaining 19.8\% on lifting the output of Z3 back to the instance model. From these percentages, we observe that the time spent by Z3 is proportionally small. This suggests that the subformulas delegated to SMT have been handled by efficient decision procedures.

We note that the proportionally high execution times of the SMT part is due to the expansion and substitution processes (Algs. \ref{algExpand} and \ref{algSubstitute}) being computationally expensive. Further, these processes may have to be executed multiple times because an instance model may pass the check in Step 3 of Fig.~\ref{fig:process}, meaning that the SMT step will be invoked, and yet the constructed SMT-LIB formula may turn out to be unsatisfiable. As a consequence, the (cumulative) execution time of the SMT part could be significant, and, as we observe in Case~A and Case~C, it could be even higher than the execution time of the search part.

As expected, Fig.~\ref{fig:RQ2} suggests an exponential trend in the execution times with the size of the instance models increasing. Despite this, all runs of our solver were successful, meaning that the solver maintained its 100\% success rate (reported in RQ1) over substantially larger instance models. In Case~A and Case~B, which, as we argued in RQ1, are less complex than Case~C, the execution times increase slowly and within reasonable range. In Case~C, one needs to take additional measures for scalability, should one need instance models that are much larger than those in our experiments. Particularly, one may utilize existing mechanisms for enhancing the performance of constraint solving, e.g., model slicing~\cite{Slicing,Slicing1}, parallelization~\cite{paraAlloy,Ranger}, and bound reduction~\cite{JordiBound}. 

\vspace*{.5em}
\begin{mdframed}[style=MyFrame]
{\it  The answer to \textbf{RQ2} is that, for our most complex case study, Case~C, PLEDGE 
generated valid data samples with over 10K objects in less than 13~hours. We believe this level of performance is practical for testing, considering the following two factors: First, one does not have to wait until testing time to initiate the generation of data; this can be done well in advance and as soon as one has a stable test data model at hand. Second, given the observed execution times of PLEDGE  over three industrial case studies, we anticipate that one would be able to generate large volumes of (logically valid) data overnight, thus minimizing the wait time by test engineers.  }\end{mdframed}

\sectopic{RQ3.}  As shown in Table~\ref{tab:RQ1} and discussed in RQ1, all the tools we have experimented with were applicable to Case~B and could quickly produce an instance model containing a single instance of the root class in that case study. RQ3 addresses a natural follow-on question: When all the tools happen to be applicable in a particular situation, which one would be the best to use for generating \emph{large} volumes of system test data? Given the results of RQ1, we elaborate RQ3 into two more detailed questions: (RQ3.1) Can one expect Alloy and UMLtoCSP to maintain their execution-time advantage over PLEDGE when they are tasked with generating larger instance models? And similarly, (RQ3.2) can PLEDGE maintain its advantage over the baseline solver as the instance models grow in size?

To answer RQ3.1 and RQ3.2, we subjected Alloy, UMLtoCSP, and the baseline solver to the same experiment as in RQ2, but restricted to Case~B only. For each of these three tools, we measured the time it took to generate valid instance models with 20, 40, 60, 80, and 100 instances of the root class (in Case~B). Similarly to RQ1, for each selected number of root classes, we set the scope parameters of Alloy and UMLtoCSP to a minimum within which they could find a solution. For example, the minimum scope that would enable Alloy to produce 40 valid instances from the root class of Case~B is 80, as all attempts with a smaller scope failed. In other words, the results we report in RQ3 for Alloy and UMLtoCSP represent the best average execution times that one can achieve, noting that in a real-world setting, practitioners would not know what the minimum scope is without experimentation. As for the baseline solver, we set the maximum number of iterations to 1000; this was sufficient for the baseline solver to succeed in all of its runs in the above-described experiment.

\begin{table*}[t]
\centering
\fontsize{7}{12}\selectfont
\setlength{\tabcolsep}{2pt}
\caption{Average Execution Times for Alloy, UMLtoCSP, the Baseline Solver and PLEDGE over Case~B}\label{tab:RQ3}
\centering
\begin{tabular}{|p{5pt}|p{2.1cm}|p{2.1cm}|p{2.1cm}|p{2.1cm}|p{2.1cm}|p{2.1cm}|}

\cline{3-7}

 \multicolumn{2}{c|}{} & \multicolumn{5}{c|}{ \centering\textbf{Number of Requested Root Class Instances}}   \\
\cline{3-7}
 \multicolumn{2}{c|}{} &   \multicolumn{1}{p{2.1cm}|}{\centering \textbf{20}}  & \multicolumn{1}{p{2.1cm}|}{ \centering \textbf{40}}& \multicolumn{1}{p{2.1cm}|}{ \centering \textbf{60}} & \multicolumn{1}{p{2.1cm}|}{\centering \textbf{80}}& \multicolumn{1}{p{2.1cm}|}{ \centering \textbf{100}}\\
\hline

 \multicolumn{1}{|p{5pt}|}{\multirow{11}{*}{{\rotatebox{90}{{\textbf{Average execution time~~~~~~~~~~~~~~~~~~~~~~~~~}
}}}}} &
  \multirow{3}{2.1cm}{\textbf{Alloy}}  &  
  \multirow{3}{2.1cm}{{0.98 min} \\ {SD = 0.05 } \\ p-value = $1.9 * 10^{-6}$} & 
   \multirow{3}{2.1cm}{{5.93 min} \\ {SD = 0.24} \\ p-value = $1.2 * 10^{-15}$} & 
   \multirow{3}{2.1cm}{{17.73 min} \\ {SD = 1.41} \\ p-value = $6.5 * 10^{-11}$} & 
   \multirow{3}{2.1cm}{{44.8 min} \\ {SD = 2.03} \\ p-value =  $1.7 * 10^{-13}$}  & 
      \multirow{3}{2.1cm}{{124.82 min} \\ {SD = 2.47} \\ p-value = $2.2 * 10^{-16}$} 
 \\
& & & & & & 
 \\
& & & & & & 
 \\
 
\cline{2-7}
 &
  \multirow{3}{2.1cm}{\textbf{UML2CSP}}  &  
  \multirow{3}{2.1cm}{{83.69 min} \\ {SD = 2.48} \\ p-value = $2.9 * 10^{-7}$} & 
   \multirow{3}{2.1cm}{Memory Overflow} & 
   \multirow{3}{2.1cm}{Memory Overflow} & 
   \multirow{3}{2.1cm}{Memory Overflow}  & 
      \multirow{3}{2.1cm}{Memory Overflow} 
 \\
& & & & & & 
 \\
& & & & & & 
 \\

\cline{2-7}
 &
  \multirow{3}{2.1cm}{\textbf{Baseline Solver}}  &  
  \multirow{3}{2.1cm}{{7.46 min} \\ {SD = 5.21} \\ p-value = $2.4 * 10^{-5}$} & 
   \multirow{3}{2.1cm}{{33.65 min} \\ {SD = 3.82} \\ p-value = $5.1 * 10^{-5}$}& 
   \multirow{3}{2.1cm}{{79.25 min} \\ {SD = 9.98} \\ p-value = $2.3 * 10^{-5}$}& 
   \multirow{3}{2.1cm}{{100.75 min} \\ {SD = 8.03} \\ p-value = $4 * 10^{-4}$}  & 
      \multirow{3}{2.1cm}{{134.93 min} \\ {SD = 16.76} \\ p-value = $1.7 * 10^{-5}$} 
 \\
& & & & & & 
 \\
& & & & & & 
 \\
\cline{2-7}
  &
  \multirow{2}{2.1cm}{\textbf{PLEDGE}}  &  
  \multirow{2}{2.1cm}{{ 0.47 min} \\ {SD = 0.12 }} & 
   \multirow{2}{2.1cm}{{1.04 min} \\ {SD = 0.25} } & 
   \multirow{2}{2.1cm}{{1.8 min} \\ {SD = 0.48} } & 
   \multirow{2}{2.1cm}{{3.13 min} \\ {SD = 0.76} }  & 
      \multirow{2}{2.1cm}{{5.58 min} \\ {SD = 1.08}} 
 \\
& & & & & & 
 \\
 
\hline

\end{tabular}
\normalsize
\vspace*{1em}
\end{table*}

Table~\ref{tab:RQ3} shows the execution times of Alloy, UMLtoCSP and the baseline solver for producing valid instance models containing 20 to 100 instances of the root class in Case~B. For easier comparison, the table further shows (in the last row) the results for PLEDGE, repeated from RQ2. To account for random variation and in a similar manner to RQ2, we calculated average execution times and their standard deviation (SD) based on ten runs. For example, Alloy took on average 124.82 min (SD = 2.47) to generate instance models with 100 instances of Case~B's root class. We note that UMLtoCSP was able to generate only instance models with up to 20 instances of the root class; all attempts with larger sizes (40, 60, 80, and 100) resulted in memory blow-up. For each selected number of root instances, Table~\ref{tab:RQ3} further presents the t-test p-values for the execution times of PLEDGE compared against those of Alloy, UMLtoCSP, and the baseline. For example, the p-value between the execution-time distribution of PLEDGE and that of Alloy, when generating 100 instances of the root class, is $2.2 * 10^{-16}$. The computed p-values indicate that the disparities between the execution times of PLEDGE and those of the considered alternatives are statistically significant. 

As seen from the table, on average, UMLtoCSP takes approximately 178 times longer to produce an instance model with 20 instances of the root class when compared to PLEDGE. As for the baseline, the average execution time is approximately 14, 32, 44, 31, and 24 times longer than that of PLEDGE for producing instance models with 20, 40, 60, 80, and 100 instances of the root class, respectively. Finally, and as for Alloy, the average execution times are consistently higher than those of PLEDGE. For instance models with up to 40 instances of the root class, Alloy performs reasonably well in comparison; however, for larger instance models, the performance difference between Alloy and PLEDGE becomes increasingly more pronounced. 
As an extra step and to examine whether the performance difference seen above could be narrowed, we experimented with an evolutionary-based version of Alloy, named EvoAlloy~\cite{EvoAlloy}. We tried various configurations for EvoAlloy over Case B (where EvoAlloy is applicable) by varying the population size and the cross-over and mutation probabilities; however, the tool was outperformed by Alloy in all the configurations that we attempted. Consequently, considering that Alloy is already included in our analysis, we chose to not include EvoAlloy in Table~\ref{tab:RQ3}, since this would not bring additional insights into the discussion.

\vspace*{.5em}
\begin{mdframed}[style=MyFrame]
{\it 
The answer to \textbf{RQ3} is that Alloy and UMLtoCSP did not maintain the performance edge they had over Case~B for larger instance models. 
UMLtoCSP produced results only for the smallest case experimented with in RQ3. Alloy was consistently outperformed by PLEDGE in all the RQ3 experiments. From a practical standpoint, Alloy scaled well up to a certain threshold, but beyond that, the performance difference between PLEDGE and Alloy became more evident. 
The baseline solver was consistently outperformed by both PLEDGE and Alloy. In short, the RQ3 results suggest that PLEDGE is the fastest among the alternatives considered for generating large volumes of system testing data.}\end{mdframed}

\section{Limitations and Threats to Validity}\label{sec:threats}
\sectopic{Limitations.}
PLEDGE is non-exhaustive. This means that, given a size bound on the instance model to build (the universe), PLEDGE can neither guarantee that it will find a solution within this bound when there is one, nor guarantee the absence of such a solution when there are none. In other words, PLEDGE forgoes completeness for better applicability and scalability of constraint solving. As a result, and more precisely speaking, PLEDGE is unable to prove bounded (un)satisfiability. This is not a major limitation in our context, since our goal is not to provide proofs of (un)satisfiability, but rather to generate test data. 
In contrast to completeness, soundness is guaranteed by PLEDGE. More precisely and as we explain below, PLEDGE is as sound as its underlying OCL evaluator, Eclipse OCL (see Section~\ref{sec:implementation}). Eclipse OCL is the de-facto OCL evaluation tool, leaving little to no room for instance models that are logically invalid to be deemed valid or vice versa. As discussed in Section~\ref{sec:approach}, PLEDGE enforces by construction the basic constraints related to class diagrams, e.g., type conformance for association ends. Further, the derived NNF constraint is logically (SAT- and UNSAT-) equivalent to the conjunction of the user and multiplicity constraints, recalling from Section~\ref{sec:ocl} that the theoretical basis for deriving NNF constraints are De Morgan's transformation rules. Consequently, when PLEDGE deems an instance model $I$ to be valid because it satisfies the NNF constraint, $I$ is indeed a valid solution for the original constraints as per the evaluation performed by Eclipse OCL.

Generally speaking, in any situation where both search and SMT are inefficient, PLEDGE will be inefficient too. One such situation (not encountered in our case studies) is when the constraints contain recursive user-defined operations. 
We inherit some limitations from the search-based solver we build on. The solver, despite covering the entire OCL, provides only coarse-grained fitness functions for certain constructs, e.g., \ocl{subOrderedSet}~\cite{Soltana:2017}. This does not pose a practical problem, since the constructs in question are rarely used.

\vspace*{.1em}

\sectopic{Threats to validity.} Internal and external validity are the validity aspects most pertinent to our empirical evaluation.

With regard to internal validity, we need to note the following: Alloy requires upfront bounds on signature (class) instantiations; and UMLtoCSP requires bounds on both class and association instantiations. Selecting these bounds is critical and yet non-trivial, noting that large bounds often lead to time and~/~or memory blow-up. In our experiments, we were as conservative as possible when comparing to Alloy and UMLtoCSP. Specifically, we minimized the bounds with hindsight from the hybrid and baseline solvers' generated data. By giving Alloy and UMLtoCSP this advantage, we mitigate as much as possible the confounding effects posed by the tuning of the bounds.

As for external validity, we note that we applied PLEDGE to three industrial case studies from different domains. 
The evaluation results are easy to interpret, and clearly suggest that PLEDGE is widely applicable and beneficial, particularly for 
complex data generation problems that are common in system testing practice. 
We also note that our three case studies were developed by different modeling teams in different domains. The team members all have background in model-driven engineering and testing, with levels of experience ranging from four to more than twenty years. The material for each case study was validated by experts from the underlying domain (legal experts for Case A, automotive experts for Case B, and satellite communications experts for Case~C). Our case study material was developed and finalized long before we started working on PLEDGE. This material was used as-is, i.e., without any changes, in our evaluation. Note that, for confidentiality reasons, the case study material we provide online has been anonymized, but from a logical standpoint, this online material is equivalent to the original material; PLEDGE yields the same results on the anonymized case studies as on the original ones. We believe that our case studies provide realistic, diverse, and unbiased examples of what PLEDGE needs to be able to handle in practice. This said, as with any type of case study research, additional case studies remain necessary for gaining further confidence in the generalizability of our approach.

\section{Related Work}\label{sec:related}
Numerous approaches exist for automated test data generation. 
Several of these approaches work by deriving test data from source code~\cite{surevyCode}. 
For example, CAUT~\cite{CAUT} uses dynamic symbolic execution and linear programming to generate unit test data for C programs. 
Given a small size bound, Korat~\cite{Korat} generates all nonisomorphic test inputs for Java programs annotated using the Java Modeling Language (JML)~\cite{JML}. SUSHI~\cite{SUSHI1,SUSHI2} combines symbolic execution and metaheuristic search to built input data that can exercise a given path in Java code.

White-box approaches like the above are usually inapplicable to system testing, which, as we noted earlier, is a primarily black-box activity~\cite{Utting:07}. In particular, white-box approaches cannot be easily used when the source code is composed of several heterogeneous languages or when the code is inaccessible / unavailable due to reasons such as confidentiality and the presence of third-party components.  These situations are common when testing at the system level. In contrast to white-box test data generation approaches, PLEDGE  requires no knowledge about the source code of the system under test, and can be applied as soon as one has a data model characterizing the system inputs.

PLEDGE  relates more closely to those that work based on a conceptual specification of data. In the context of UML, most such approaches utilize either only Class Diagrams~\cite{Dania:2016,Shaukat:2013,Shaukat:2016}, or a combination of Class Diagrams and behavioral models such as State Machines~\cite{behave1} and Sequence Diagrams~\cite{behave2}. Regardless of the exact formulation of UML-based data generation, a core challenge in this line of work is the satisfaction of OCL constraints~\cite{Gonzalez:2014}. The most commonly used strategy for OCL solving is through translation to SAT, SMT, and other formalisms equipped with solvers. Alloy~\cite{JacksonBook2012} and UMLtoCSP~\cite{UMLtoCSP} are two notable approaches in this class. Alloy is a general-purpose SAT-based analyzer for first-order logic. Alloy can act as an OCL solver when complemented with a tool such as UML2Alloy~\cite{%AnastasakisBGR10,
UML2Alloy}, which adapts UML class diagrams and OCL constraints to Alloy's input language. UMLtoCSP, which is specifically geared toward OCL solving, has as translation target a Prolog-based constraint programming language. Since Alloy and UMLtoCSP are currently the main technologies used for OCL solving, we empirically compared PLEDGE  against them in our evaluation (Section~\ref{sec:evaluation}). In addition to Alloy and UMLtoCSP, there are several threads of work where OCL constraints are solved through translation, with the main translation targets being SAT~\cite{SoekenWD11,KuhlmannG12,PrzigodaSWD16} and SMT~\cite{CantenotAB14,Przigoda:2016,Dania:2016}.

The above approaches have two main limitations: First, they all have to translate class diagrams and OCL constraints into other formalisms. Since the formalisms considered are not sufficiently expressive, compromises have had to be made in terms of supported features. For example, we faced applicability problems with Alloy and UMLtoCSP in two out of our three case studies. Another limitation is that these approaches do not scale well for test data generation; we empirically demonstrated this for Alloy and UMLtoCSP. 

To mitigate challenges posed by undecidability and the of lack scalability, there has been work on identifying decidable fragments in OCL~\cite{OCL-Lite,simp1, simp2}. In practice, narrowing OCL to decidable fragments is problematic, since doing so often makes certain constraints inexpressible, and further poses a usability challenge to users, even when the constraints \hbox{can be written in a decidable fragment.}

An alternative strategy for OCL solving is metaheuristic search~\cite{Shaukat:2013,Shaukat:2016,Soltana:2017,Hassan}.
Due to their non-exhaustive nature, search-based approaches cannot provide guarantees about either satisfiability or unsatisfiability. Nevertheless, for activities such as simulation and testing, such guarantees are not necessary, noting that the objective of OCL solving here is not to assess the quality of a model, but rather generate synthetic data from a model that has been validated a priori. Search-based OCL solving approaches are parallelizable, and have been shown to scale well for large and complex data generation problems. Further, these approaches cover nearly the entire OCL language -- a direct benefit of not requiring a translation to another formalism -- and thus present a major applicability advantage. Despite these advantages, search is a randomized process and not as effective as a decision procedure for handling the decidable fragments of OCL.

Metaheuristic search has also been considered for solving Alloy specifications. Notably, using Alloy's intermediate relational models as seeds, EvoAlloy provides a population-based evolutionary algorithm aimed at improving the success rate of solving large Alloy specifications~\cite{EvoAlloy}. EvoAlloy, by virtue of being based on Alloy's specification language, inherits the expressiveness restrictions and applicability issues of Alloy. In the context of our work and as discussed in Section~\ref{sec:results}, EvoAlloy is applicable to only one of our case studies (Case~B).

More broadly, the Search-Based Software Testing (SBST) community has long studied metaheuristic search for generating test data. Nevertheless, aside from the OCL- and Alloy-related work discussed above, the existing literature on search-based test data generation is oriented primarily around software code~\cite{McMinn:04,AliBHP10,HarmanMZ12}. For example, Harman and McMinn~\cite{HarmanM10} combine evolutionary (global) and Hill Climbing (local) search techniques for generating test data from Java programs while maximizing branch coverage. 
Using control-dependency graphs built from C code, Pargas et al.~\cite{pargas1999test} devise a genetic algorithm for test case generation with the goal of maximizing statement and branch coverage. 
Panichella et al.~\cite{Annibale} investigate multi- and many-objective optimization for building Java test suites that simultaneously  maximize several coverage criteria, including statement, branch, and strong mutation coverage. All the above-cited approaches have some form of constraint solving embedded in them. The constraints targeted by these approaches are nevertheless at the source-code level, thus making them either inapplicable or unsuitable for system-level constraints, which are typically written in high-level languages \hbox{based on first-order logic such as OCL.} 

While PLEDGE cannot be compared directly with approaches for code-level constraint solving, the hybridization idea pursued by PLEDGE is analogous to some existing work in that area. For example, Inkumsah et al.~\cite{nkumsahX08} combine search and Dynamic Symbolic Execution (DSE) for improving scalability in the structural testing of object-oriented systems. Here, search and DSE are combined in a sequential manner (first search followed by DSE), the assumption being that the constraints handled by each technique are independent from one another. In our context, and as we discussed in Section~\ref{sec:share}, one cannot always separate OCL subformulas in a way that search and SMT can be applied to them sequentially and yet effectively. This is why PLEDGE invokes search and SMT in an alternating and iterative manner, thus allowing the two techniques to work in tandem rather than sequentially.

There is work outside UML on specification-based data generation. A recent strand exploits a combination of exhaustive search and graph transformation for generating well-formed graph structures~\cite{VarroICSE}. Here, the underlying constraints are defined through graph-logic expressions. While this approach has the advantage of providing guarantees in terms of satisfiability,  the underlying  constraint language has limited expressiveness and covers only a small fragment of the OCL metamodel. Furthermore, the approach does not provide adequate support for integer, real, and string variables beyond treating them as enumerations. PLEDGE  covers the entire OCL and does not suffer from the above limitations.

Another data generation approach from outside the UML community is TestBlox~\cite{TestBlox}. TestBlox aims to produce large, valid, and statistically representative data sets from multidimensional models to test the performance of big-data computing platforms. This approach admits a mix of hard (dimension or integrity) and soft (statistical) constraints. The approach relies on a customized solver that is able to generate, for certain types of constraints, test data in polynomial time with respect to the size of data instances~\cite{TestBlox}. This low computational complexity nonetheless comes at the expense of some strong restrictions. First, TestBlox requires that the generated data should be a DAG (Directed Acyclic Graph). Second, TestBlox is meant at solving only certain constraints, all of which have to do with the structural aspects of the generated data, e.g., cardinalities and the direction of links between objects. The result is that, in terms of expressiveness, TestBlox's declarative modeling language covers only a fragment of OCL.

Although PLEDGE cannot offer the same level of scalability as TestBlox, it is more practical to use in the context of system testing. PLEDGE does not restrict the structure of the instance models. In addition, it supports the entire OCL and thus a much larger spectrum of constraints, including constraints over primitive attributes. One can of course foresee hybrid approaches where TestBlox is employed as an alternative or complement to the search component of PLEDGE. However, for the reasons discussed above, TestBlox per se is not an alternative to PLEDGE for system-level test data generation purposes. For example, TestBlox cannot be applied to any of the three case studies used in Section~\ref{sec:evaluation} for evaluating PLEDGE. This is because these case studies all involve constraints, e.g., over attributes, that are not expressible in TestBlox.

In relation to soft (statistical) constraints supported by TestBlox, we note the following: We have developed in our previous work an approach for guiding OCL solving toward generating statistically representative data~\cite{Soltana:2017}. That approach works by continuously monitoring the data produced by an OCL solver and dynamically adding ``corrective'' soft constraints to align the properties of the data with a given set of statistical distributions. PLEDGE can be readily applied for generating statistically representative data. Evaluating PLEDGE in such a usage scenario is nevertheless outside the scope of this article. This said, we show in this article that PLEDGE outperforms the baseline solver from our previous work which was used for generating statistically representative data. 
We thus anticipate PLEDGE to also outperform the baseline when used in a similar manner.

\section{Conclusion}\label{sec:conclusion}
We proposed a tool-supported approach based on UML for generating system test data under constraints. Our approach hybridizes metaheuristic search and SMT for solving complex constraints expressed in OCL. To our knowledge, we are the first to develop such a hybrid approach for constraint solving. We evaluated our approach on three industrial case studies, showing that the approach brings about important benefits in terms of applicability and scalability, when used on complex real-world problems.

Our current approach uses SMT only for solving constraints over primitive attributes. In the future, we would like to look into ways to exploit SMT more broadly, and thus further tap into the potential of background SMT theories. Another topic for future work has to do with how we delegate solving tasks to search and SMT. Currently, our delegation process is static. We plan to investigate whether a dynamic delegation strategy, e.g., using a machine-learning-based feedback loop, can provide advantages.

\bibliographystyle{ACM-Reference-Format-Journals}
\bibliography{paper}

\newpage

\appendix
\noindent{\LARGE Appendix}
\section{Rules for Delegating Solving Tasks to Search and SMT}\label{sec:rules}
\begin{figure}[!h]
	\centering
	\includegraphics[ scale=0.24,angle = 90]{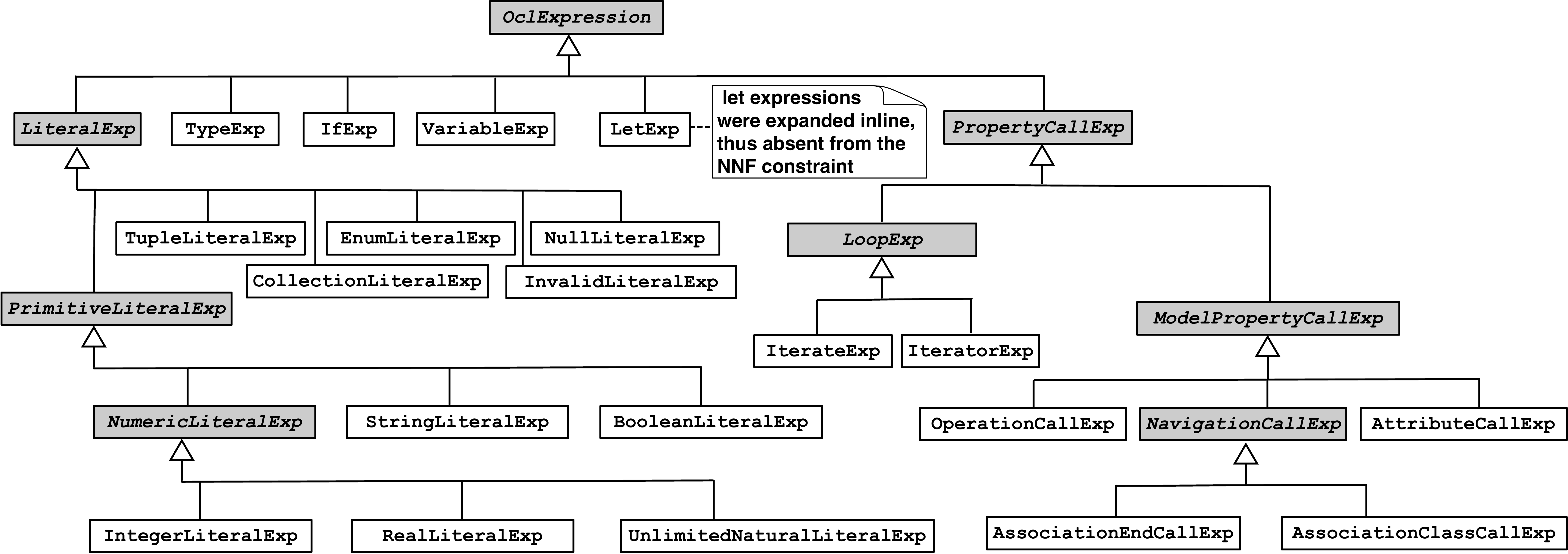}
	\caption{Simplified View of the OCL 2.4 Metamodel\label{fig:meta}}
	\vspace*{-2em}
\end{figure}

Table~\ref{tab:labels} provides our rule set for labeling the AST nodes of OCL expressions.
The rules cover \emph{all} the AST node types in the OCL 2.4 metamodel (\url{https://www.omg.org/spec/OCL/2.4/}). Fig.~\ref{fig:meta} depicts a simplified version of this metamodel, with only the inheritance (and not the composition) relationships shown. In the figure, \hbox{abstract AST node types are shaded grey.}
\textcolor{black}{Given an AST (for an NNF constraint), our labeling procedure visits all the nodes in the AST and assigns labels to all the visited nodes except those representing constants. Our comprehensive labeling of the AST ensures that we always come down with a decision about whether to delegate a specific solving task to search, SMT, or both. Whether the decisions made by our labeling procedure lead to an effective solving of the constraints at hand is a separate question, which we investigated empirically in Section~\ref{sec:evaluation}.}

 \pagebreak 

Each row in Table~\ref{tab:labels} defines one labeling rule. 
The first column of the table indicates the node type (from the metamodel of  Fig.~\ref{fig:meta}). When the different subtypes of the node type in the first column require different treatments (i.e., need to be assigned different labels), the second column enumerates the subtypes separately. The fourth column shows the label to attach to a node type (or subtype) if the condition specified in the third column holds. The rationale behind the label attached to a given node (sub)type is provided alongside the condition in the third row. We note that, when a given node type has multiple conditions, $C_1,\ldots,C_n$, an if-then-else semantics is intended, i.e.,
$\textbf{if}\ (C_1)\ \textbf{then}\ \textit{label}_1\ \textbf{else if}\ (C_2)\ \textbf{then}\ \textit{label}_2 \ldots \textbf{else if}\ (C_n)\ \textbf{then}\ \textit{label}_n$. Here, the last condition, $C_n$, is always the default (catch-all) condition.

\begin{small}
\setlength\tabcolsep{2pt}
\begin{longtable}[h]{|p{13em}|p{12em}|p{15em}|p{3em}|}
\caption{Complete List of Rules for Labeling OCL AST Nodes~\label{tab:labels} (the Subtypes column is filled only when the labeling is not uniform across subtypes)}\\
\cline{1-4} 
\centering \textbf{Node type} & \multicolumn{1}{c|}{  \textbf{Subtypes}}&   \multicolumn{1}{c|}{\color{black}\textbf{Condition \& Rationale}} &   \multicolumn{1}{c|}{ \textbf{Label}}  \\ 
\cline{1-4} 
\endfirsthead
\endhead
\ocl{PrimitiveLiteralExp}'s \newline subtypes, \ocl{EnumLiteralExp}  & - & 
\multirow{6}{15em}{\textbf{Condition}: - \\ 
\textbf{Rationale}: \textcolor{black}{The targeted nodes represent constants such as numeric and string literals. Constants are left unlabelled because they per se do not induce any solving tasks.}}
 & - \\
 & & & \\ 
 & & & \\
 & & & \\
 & & & \\
\cline{1-4}

 \ocl{InvalidLiteralExp}, \ocl{TypeExp} & -  & 
\multirow{7}{15em}{\textbf{Condition}: - \\ 
\textbf{Rationale}:  SMT does not support OCL's invalid and type expressions. Such expressions, which are mostly used to constrain the structure of the instance model, are always assigned to search.  }
 & \textit{search} \\
 & & & \\ 
 & & & \\
 & & & \\
 & & & \\
 & & & \\
  & & & \\
\cline{1-4} 

\ocl{LoopExp}'s subtypes  & -  & 
\multirow{6}{15em}{\textbf{Condition}: - \\ 
\textbf{Rationale}: As discussed in Section~\ref{sec:background}, SMT is not our preferred alternative for handling OCL's loop expressions, including quantifiers. We therefore delegate all loop expressions to search.}
 & \textit{search} \\
 & & & \\ 
 & & & \\
 & & & \\
 & & & \\
 & & & \\
 & & & \\
\cline{1-4} 

\ocl{CollectionLiteralExp}, \ocl{TupleLiteralExp} & -  & 
\multirow{7}{15em}{\textbf{Condition}: - \\ 
\textbf{Rationale}:  Similar to loop experssions, collections and tuple are not readily matched to efficient SMT decision procedure (see Section~\ref{sec:approach}). Collections and tuple expressions are thus always assigned to search.  }
 & \textit{search} \\
 & & & \\ 
 & & & \\
 & & & \\
 & & & \\
 & & & \\
\cline{1-4} 

\ocl{NavigationCallExp}'s subtypes & -
 & 
 \multirow{7}{15em}{\textbf{Condition}: - \\ 
\textbf{Rationale}: Evaluating a navigation expression yields one of the following: an object, a collection, or an invalid / null. None of these return types are primitive. As discussed in Section~\ref{sec:approach}, SMT's involvement is narrowed to subformulas containing primitive types only. We thus always assign navigations to search.}
   & \textit{search}  \\
    & & & \\ 
 & & & \\
 & & & \\
 & & & \\
 & & & \\
 & & & \\
 & & & \\

  & & & \\
  & & & \\
\cline{1-4} 
\hline

\multirow{1}{*}{\ocl{VariableExp}}, \multirow{1}{*}{\ocl{AttributeCallExp}}& - & 
\multirow{3}{15em}{\textbf{Condition 1}: If the attribute / variable is non-primitive. \\ 
\textbf{Rationale}:   When the accessed attribute / variable represents}
  & \textit{search}\\ 
    & & & \\
    & & & \\

        & &   an object, a collection, or an invalid / null value, the node is labeled search. The rationale is the same as that for the \ocl{NavigationCallExp} subtypes in the previous row of the table.& \\
  \cline{3-4}
 &  &
 \multirow{7}{15em}{\textbf{Condition~2}:    If the attribute / variable is primitive, but is contained  directly or indirectly within the body of an \ocl{exists}, \ocl{select}, \ocl{reject}, \ocl{any}, \ocl{isUnique}, or \ocl{one}. 
 \\ \textbf{Rationale}:  In principle, nodes representing primitive attributes or variables should be delegated to SMT.  This rule is an exception to the principle, allowing both SMT and search to alter the value of certain attributes / variables. For example, consider the subformula \ocl{c->select(a = 3)->notEmpty()}. Although \ocl{a} is a primitive attribute, search still needs visibility into it. This is because \ocl{a} appears within a \ocl{select}, which is to be handled by search. If search is barred from manipulating \ocl{a}, then search would be extremely unlikely to be able to evolve the structure of the instance model to a point where SMT can produce a satisfying assignment. The same explanation applies to \ocl{exists}, \ocl{reject}, \ocl{any}, \ocl{isUnique}, and \ocl{one}, all of which are to be handled by search.} 
  & \textit{both}\\ 
      & & & \\ 
 & & & \\
 & & & \\
 & & & \\
 & & & \\
 & & & \\
 & & & \\
 & & & \\
 & & & \\
  & & & \\
  & & & \\
    & & & \\
 & & & \\
 & & & \\
  & & & \\
  & & & \\
    & & & \\
    & & & \\
 & & & \\
  & & & \\
  & & & \\
    & & & \\
     & & & \\
  & & & \\
  & & & \\
    & & & \\

  \cline{3-4}
      & &
       \multirow{7}{15em}{\textbf{Condition~3}: If neither Condition 1 nor Condition~2 holds. \\ 
       \textbf{Rationale}:  Here, we are in the situation where the node in question represents an attribute / variable meeting both of the following criteria: (1) the node is primitive, and (2) the node is not contained in \ocl{exists}, \ocl{select}, \ocl{reject}, \ocl{any}, \ocl{isUnique}, or \ocl{one}. In this situation, SMT can find a satisfying assingment independently of search. We thus assign the node to SMT.} 
  & \textit{SMT} \\ 
       & & & \\
  & & & \\
  & & & \\
  & & & \\
  & & & \\
  & & & \\
    & & & \\
    & & & \\
      & & & \\
 & & & \\
  & & & \\
  & & & \\
      
  \cline{1-4}
      
\multirow{1}{*}{\ocl{IfExp}} & - &
 \multirow{2}{15em}{\textbf{Condition 1}: If the body of the if-then-else has a non-primitive   
 } 
& \textit{search}\\
 & & & \\

\multirow{1}{*}{} &  &
 \multirow{3}{15em}{return type. \\ 
 \textbf{Rationale}: Same rationale as that given for Condition 1 of the \ocl{VariableExp} and \ocl{AttributeCallExp} types.
 } 
& \\
 & & & \\
  & & & \\
   & & & \\
    & & & \\
\cline{3-4}

 & &
 \multirow{1}{15em}{\textbf{Condition~2}:    If the the body of the if-then-else has a primitive return type, but is directly or indirectly contained within an \ocl{exists}, \ocl{select}, \ocl{reject}, \ocl{any}, \ocl{isUnique}, or \ocl{one}.\\
  \textbf{Rationale}:  Same rationale as that given for Condition~2 of the \ocl{VariableExp} and \ocl{AttributeCallExp} types. }
& \textit{both}  \\
   & &   \newline

   & \\
     & & &\\
   & & &\\
     & & &\\
   & & &\\
      & & &\\
  \cline{3-4}
  &  & \multirow{6}{15em}{\textbf{Condition~3}:     If neither Condition 1 nor Condition~2 holds.  \\ 
 \textbf{Rationale}: The rationale is the same as that for Condition~3 of the \ocl{VariableExp} and \ocl{AttributeCallExp} types.
} 
 & \textit{SMT}\\ 
  & & & \\
  & & &\\
   & & &\\
     & & & \\
  & & &\\
\cline{1-4}

\cline{1-4} 
 \multirow{1}{*}{\ocl{OperationCallExp}}  & \multirow{1}{10em}{\ocl{and}, \ocl{or}, \ocl{not}, \ocl{<}, \ocl{>}, \ocl{<=}, \ocl{>=}, \ocl{=}, \ocl{<>}, \ocl{+}, \ocl{-}, \ocl{*}, \ocl{/}%, \ocl{\^{}}
 } & 
  \multirow{7}{15em}{\textbf{Condition 1}:    If all children are unlabeled.   \\ 
 \textbf{Rationale}:  The labels of these operation nodes are deduced from those of their children. This particular condition covers the situation where all the children are unlabelled. For example, the node representing \ocl{or} in the following subformula is left unlabeled: \ocl{false or true}. This is because the children (\ocl{false} and \ocl{true}) are constants and unlabeled. Neither search nor SMT should attempt to solve an unlabeled subformula.} 
 & - \\ 
    & & &\\
     & & & \\
        & & &\\
              & & &\\
     & & & \\
        & & &\\
           & & &\\
     & & & \\
           & & &\\
     & & & \\
     & & & \\
        & & &\\
     & & & \\
        & & &\\
 \cline{3-4}
  & &   
    \multirow{7}{15em}{\textbf{Condition~2}:      If all labeled children have \textit{search} as label.  \\ 
 \textbf{Rationale}: Since all the (labeled) children are delegated to search, intervention from SMT is unwarranted; the whole subformula is therefore to be handled by search.   } 
 & \textit{search} \\ 
         & & &\\
     & & & \\
        & & &\\
     & & & \\
             & & &\\
     & & & \\
  \cline{3-4}
    & &  \multirow{7}{15em}{\textbf{Condition~3}:   If all labeled children have \textit{SMT} as label.  \\ 
 \textbf{Rationale}: Since all the (labeled) children are delegated to SMT, intervention from search is unwarranted; the whole subformula is therefore to be handled by SMT.} 
 & \textit{SMT} \\ 
            & & &\\
     & & & \\
        & & &\\
     & & & \\
          & & & \\
             & & &\\
         \cline{3-4}

        & & 
 \multirow{2}{15em}{\textbf{Condition~4}:  If all labeled children have \textit{both} as label.  
} 
 & \textit{both} \\ 
            & & &\\

         & & 
 \multirow{2}{15em}{\textbf{Rationale}: Since all the (labeled) children require both search and SMT, so does the parent node.  
} 
 & \\ 
            & & &\\
   & & &\\
               \cline{3-4}

               & &   \multirow{7}{15em}{\textbf{Condition~5}:   If none of the above four conditions hold. \\ 
 \textbf{Rationale}:  Here, the (labeled) children include at least two (and potentially all three) possible labels: \textit{search}, \textit{SMT}, and \textit{both}. This means that, just like in Condition~4, the intervention of both search and SMT is warranted to solve the underlying subformula. As a result, the label assigned to the node is \textit{both}.    } 
 & \textit{both} \\ 
      & & & \\
        & & &\\
                             & & &\\
             & & & \\
             & & &\\
                     & & &\\
             & & & \\    
                     & & &\\
       & &  &\\
       
       & & & \\

\cline{2-4}

 & \multirow{1}{*}{\ocl{size}, \ocl{max}, \ocl{min}} &

    \multirow{2}{15em}{\textbf{Condition 1}:     If the operation is called  over a collection.      \\ 
 \textbf{Rationale}: The operations in this sub-type can be invoked over both collections and primitive attributes. For example, one can use the \ocl{size} operation over either a string attribute or over a set. This rule thus needs to first check the type of the element to which the operation is applied. If the calling element is a collection, then the whole subformula is delegated to search. The rationale is the same as that for \ocl{CollectionLiteralExp} and \ocl{TupleLiteralExp}, presented earlier.} 

  & \textit{search} \\ 
   & & &\\
             & & & \\
             & & &\\
                     & & &\\
             & & & \\    
                     & & &\\
                     & & &\\
             & & & \\
             & & &\\
                     & & &\\
             & & & \\    
             & & & \\    
             & & & \\    

                     & & &\\
   \cline{3-4}
         & & 
         
             \multirow{2}{15em}{\textbf{Condition~2}:  If the operation is called  over a primitive type, and at the same time, the operation is directly or indirectly contained within the body of an \ocl{exists}, \ocl{select}, \ocl{reject}, \ocl{any}, \ocl{isUnique}, or \ocl{one}.      \\ 
 \textbf{Rationale}:  Same rationale as that given for Condition~2 of the \ocl{VariableExp} and \ocl{AttributeCallExp} types.
 } 
  & \textit{both} \\ 

            & & & \\    
            & & & \\    
                     & & &\\
             & & & \\
             & & &\\
                     & & &\\
             & & & \\    
                     & & &\\
           & & & \\    
                     & & &\\

   \cline{3-4}
         & & 
            \multirow{2}{15em}{\textbf{Condition~3}:  If neither Condition 1 nor Condition 2 holds.   \\ 
 \textbf{Rationale}: This rule covers the situation where: (1) the node represents an operation that  is applied over a primitive attribute / variable, and (2)  the underlying subformula can be solved independently of search. We therefore label the node as \textit{SMT}.  } 
& \textit{SMT} \\ 

        & & & \\
             & & &\\
                     & & &\\
             & & & \\    
                    & & & \\
             & & &\\
                        & & & \\    
                    & & & \\
             & & &\\
\cline{2-4}\\[-1.5em]

\cline{2-4}
 & \multirow{1}{10em}{\ocl{concat}, \ocl{substring}, \ocl{toInteger}, \ocl{toReal}, \ocl{round}, \ocl{ceil}, \ocl{abs}, \ocl{floor}, \ocl{div}, \ocl{mod}}  & 
 
          \multirow{7}{15em}{\textbf{Condition 1}:  If the operation in question is directly or indirectly contained within the body of an \ocl{exists},  \ocl{select}, \ocl{reject}, \ocl{any}, \ocl{isUnique}, or \ocl{one}.  \\ 
 \textbf{Rationale}: All the operations here are applicable to primitive attributes / variables only. The rationale is the same as that for Condition~2 of the \ocl{VariableExp} and \ocl{AttributeCallExp} types. } 
  & \textit{both} \\ 
  & & & \\
 & & & \\
   & & & \\
 & & & \\
   & & & \\
 & & & \\
   & & & \\
 & & & \\
    & & & \\
 & & & \\

  \cline{3-4}
         & &   \multirow{2}{15em}{\textbf{Condition 2}: If Condition~1 does not hold.  \\ 
 \textbf{Rationale}:  The rationale is the same as that for Condition~2 of the \ocl{VariableExp} and \ocl{AttributeCallExp} types.}  & \textit{SMT} \\ 
              & & &\\
                     & & &\\
             & & & \\    
                    & & & \\
                      & & & \\
 \cline{2-4}

\cline{2-4}

 & \multirow{4}{10em}{\ocl{toLower}, \ocl{sum}, \ocl{toUpper}, \ocl{union}, \ocl{allInstances}, \ocl{oclIsInvalid}, \ocl{oclIsKindOf}, \ocl{oclIsTypeOf}, \ocl{oclIsUndefined}, \ocl{oclAsType}, \ocl{any}, \ocl{asBag}, \ocl{collect}, \ocl{asOrderedSet}, \ocl{asSequence}, \ocl{asSet},  \ocl{collectNested},  \ocl{count}, \ocl{excludes},  \ocl{excludesAll}, \ocl{excluding},  \ocl{flatten}, \ocl{includes},  \ocl{includesAll}, \ocl{including},  \ocl{isEmpty}, \ocl{isUnique}, \ocl{notEmpty}, \ocl{one}, \ocl{product}, \ocl{reject}, \ocl{select}, \ocl{sortedBy}, \ocl{append}, \ocl{at}, \ocl{first}, \ocl{last}, \ocl{indexOf}, \ocl{insertAt}, \ocl{prepend}, \ocl{subsequence}, \ocl{intersection}, \ocl{subOrderedSet}, \ocl{symmetricDifference} } & 
 
    \multirow{2}{15em}{\textbf{Condition}: -  \\ 
 \textbf{Rationale}: The operations in this subtype are all labeled \textit{search}. The reason is that each of these operations falls under one of the following two categories for which SMT is not an option: (1) the operation is invoked over objects, collections, and other non-primitive data types, or (2)~the operation applies to primitive types, but does not have suitable counterpart functions in SMT-LIB as discussed in Section~\ref{sec:z3}.} 

 & \textit{search} \\ 

  &  & &        \\ 
  &  & &            \\ 
      &  & &         \\ 
        &  & &        \\ 
  &  & &            \\ 
      &  & &         \\ 
        &  & &        \\ 
  &  & &            \\ 
      &  & &         \\ 
        &  & &        \\ 
  &  & &            \\ 
      &  & &         \\ 
      
            &  & &         \\ 
        &  & &        \\ 
  &  & &            \\ 
      &  & &         \\ 
            &  & &         \\ 
        &  & &        \\ 
  &  & &            \\ 
      &  & &         \\ 
              &  & &        \\ 
\cline{2-4}

 & \multirow{1}{*}{\ocl{user-defined}} &
     \multirow{2}{15em}{\textbf{Condition 1}:  If the operation is both non-recursive and further has a primitive return type.  \\ 
 \textbf{Rationale}: The rationale is the same as that for Condition~3 of the \ocl{VariableExp} and \ocl{AttributeCallExp} types. We note that this rule excludes recursive user-defined operations from being handled by SMT, even when such operations have primitive return types. This is because we } 
 & \textit{SMT} \\ 
 
            &  & &         \\ 
        &  & &        \\ 
  &  & &            \\ 
      &  & &         \\ 
            &  & &         \\ 
        &  & &        \\ 
  &  & &            \\ 
      &  & &         \\ 
              &  & &        \\ 
            &  & &         \\

 &  &
     \multirow{2}{15em}{found Z3 (the SMT solver used by PLEDGE) to be inefficient over recursive functions.} 
 &  \\ 
        &  & &        \\ 
            &  & &         \\

  \cline{3-4}
    & &  
         \multirow{2}{15em}{\textbf{Condition 2}:      If Condition~1 does not hold.    \\ 
 \textbf{Rationale}: All remaining types of user-defined operations, i.e., recursive ones and those with non-primitive return types, are assigned to search. This is because: (1)~as explained in Condition~1 above, SMT is not ideal for recursive functions, and (2) following the same rationale as that for Condition~1 of the \ocl{VariableExp} and \ocl{AttributeCallExp} types, we want only search to handle non-primitive constructs.} 
    
 & \textit{search} \\ 
              &  & &        \\ 
              &  & &        \\ 
              &  & &        \\ 
              &  & &        \\ 
              &  & &        \\ 
              
            &  & &         \\ 
        &  & &        \\ 
  &  & &            \\ 
            &  & &         \\ 
              &  & &        \\ 
              
            &  & &         \\ 
             &  & &         \\ 
               &  & &         \\ 

\cline{1-4} 

\end{longtable}
\end{small}

\section{Expanding OCL's Built-in Quantification Shortcuts during the Construction of an SMT-LIB Formula\label{sec:expand}}
Table~\ref{tab:expand} lists all the OCL constructs that involve implicit quantification and thus need to be expanded over a given instance model when an SMT-LIB formula is being built.
The first column of the table presents the OCL construct to be expanded. The second column describes how the expansion is performed when the construct is applied to non-empty collections. The third column provides the literal that replaces the construct (1) when the construct is called over an empty collection, or (2) in cases where the construct compares two collections, when at least one collection is empty.

\begin{small}
\setlength\tabcolsep{1pt}
\begin{longtable}[t!]{|p{10em}|p{20em}|p{10em}|}
\caption{Full List of OCL Constructs with Implicit Quantification\label{tab:expand}}\\
\hline 
\multicolumn{1}{|c|}{\multirow{2}{10em}{ \centering\textbf{OCL construct}}} & \multicolumn{1}{c|}{ \multirow{2}{20em}{ \centering\textbf{Description of the expansion process}}}& \multicolumn{1}{c|}{ \multirow{2}{10em}{\centering\textbf{Expansion involving empty collections}}} \\ 
\multicolumn{1}{|c|}{\multirow{1}{10em}{ }} & & \\
\hline
\endfirsthead
\endhead

\ocl{excludes} & Expanding \ocl{c->excludes(x)}, where \ocl{c} is a collection and \ocl{x} is a given element, is equivalent to expanding: \ocl{(not c->exists(i| i = x))}. & \ocl{true}\\
\hline
\ocl{excludesAll} & 
Expanding \ocl{c1->excludesAll(c2)}, where \ocl{c1} and \ocl{c2} are two collections, is equivalent to expanding: \newline \ocl{(c2->forAll(i| not c1->exists(j| i = j))}.
& \ocl{true}\\
\hline
\ocl{includes} & 
Expanding \ocl{c->includes(x)}, where \ocl{c} is a collection and \ocl{x} is a given element, is equivalent to expanding: \ocl{(c->exists(i| i = x))}. 
& \ocl{false}\\
\hline
\ocl{includesAll} & 
Expanding \ocl{c1->includesAll(c2)}, where \ocl{c1} and \ocl{c2} are two collections, is equivalent to expanding: \newline \ocl{(c2->forAll(j| c1->exists(i| i = j)))}.
& \ocl{false} \\
\hline
\ocl{isUnique} &
Expanding \ocl{c->isUnique(\variable{condition over c})}, where \ocl{c} is a collection, is equivalent to expanding: \newline \ocl{(c->forAll(i| c->forAll(j| i = j or \variable{condition over c$_i$} <> \variable{condition over c$_j$})))}.
 &\ocl{true} \\
\hline
\ocl{one} &
Expanding \ocl{c->one(\variable{condition over c})}, where \ocl{c} is a collection, is equivalent to expanding: \newline \ocl{c->exists(i| \variable{condition over c$_i$} and c->forAll(j| i <> j and not \variable{condition over c$_j$}))}.
 &\ocl{false} \\
\hline
\ocl{=}
(for sequences) & 
Expanding \ocl{s1 = s2}, where \ocl{s1} and \ocl{s2} are two sequences, is equivalent to expanding: \newline \ocl{(s1->size() = s2->size() 
and s1->asSet() ->asSequence()->forAll(i| s2->asSet()->asSequence()->at(s1->asSet() ->asSequence()->indexOf(i)) = i))}.
& If both collections are empty then \ocl{true}, otherwise \ocl{false}  \\
\hline
\ocl{<>}
(for sequences) & 
Expanding \ocl{s1 <> s2}, where \ocl{s1} and \ocl{s2} are two sequences, is equivalent to expanding: \newline \ocl{(s1->size() <> s2->size() 
or s1->asSet() ->asSequence()->exists(i| s2->asSet()->asSequence()->at(s1->asSet() ->asSequence()->indexOf(i)) <> i))}.
& If both collections are empty then \ocl{false}, otherwise \ocl{true} \\
\hline
\ocl{=}
(for bags) & 
Expanding \ocl{b1 = b2}, where \ocl{b1} and \ocl{b2} are two bags, is equivalent to expanding: \ocl{(b1->{forAll}(i| b2->exists(j| i = j and b1->count(i) = b2->count(j)))).}
& If both collections are empty then \ocl{true}, otherwise \ocl{false}  \\
\hline
\ocl{<>}
(for bags) & 
Expanding \ocl{b1 <> b2}, where \ocl{b1} and \ocl{b2} are two bags, is equivalent to expanding: \ocl{(b1->{exists}(i| b2->forAll(j| i <> j or b1->count(i) <> b2->count(j)))).}
& If both collection are empty then \ocl{false}, otherwise \ocl{true} \\
\hline
\ocl{=}
(for sets) &
Expanding \ocl{s1 = s2}, where \ocl{s1} and \ocl{s2} are two sets, is equivalent to expanding: \ocl{(s1->{forAll}(i\,| s2->{exists}(j\,| i = j)) {and} s2->{forAll}(i\,| s1->{exists}(j\,| i = j)))}
 & If both collections are empty then \ocl{true}, otherwise \ocl{false} \\
\hline
\ocl{<>}
(for sets) & 
Expanding \ocl{s1 <> s2}, where \ocl{s1} and \ocl{s2} are two sets, is equivalent to expanding: \ocl{(s1->{exists}(i\,| s2->{forAll}(j\,| i <> j)) {or} s2->{exists}(i\,| s1->{forAll}(j\,| i <> j)))}.
& If both collection are empty then \ocl{false}, otherwise \ocl{true} \\
\hline
\ocl{=}
(for ordered sets) & 
Expanding \ocl{os1 = os2}, where \ocl{os1} and \ocl{os2} are two ordered sets, is equivalent to expanding: \ocl{(os1->size() = os2->size() and os1->forAll(i\,| os2->exists(j\.| i = j)))}.
&If both collections are empty then \ocl{true}, otherwise \ocl{false}  \\
\hline
\ocl{<>}
(for ordered sets) & 
Expanding \ocl{os1 <> os2}, where \ocl{os1} and \ocl{os2} are two ordered sets, is equivalent to expanding: \ocl{(os1->size() <> os2->size() or os1->exists(i\,| os2->forAll(j\.| i <> j)))}.
& If both collection are empty then \ocl{false}, otherwise \ocl{true} \\
\hline
\ocl{=}
(between an ordered Set and a set) & 
Expanding \ocl{os = s}, where \ocl{os} is an ordered set and \ocl{s} is a set, is equivalent to expanding: \ocl{(os->asSet()->{forAll}(i\,| s->{exists}(j\,| i = j)) {and} s->{forAll}(i\,| os->asSet->{exists}(j\,| i = j)))}
& If both collections are empty then \ocl{true}, otherwise \ocl{false}  \\
\hline
\ocl{<>}
(between an ordered set and a set) &
Expanding \ocl{os <> s}, where \ocl{os} is an ordered set and \ocl{s} is a set, is equivalent to expanding: \ocl{(os->asSet()->{exists}(i\,| s->{forAll}(j\,| i <> j)) or s->{exists}(i\,| os->asSet->{forAll}(j\,| i <> j)))}
 & If both collection are empty then \ocl{false}, otherwise \ocl{true} \\
\hline
\end{longtable}
\end{small}

\section{Rules for Translating OCL to SMT-LIB}\label{sec:toSMT-LIB}
Table~\ref{tab:rules} provides the rules that we use for translating NNF OCL expressions (after the execution of the expansion and substitution processes) to SMT-LIB. 
The first column of the table indicates the type of the traversed node (from the metamodel of Fig.~\ref{fig:meta}).
The second column shows the general shape of the OCL expression represented by the node in the first column.
\hbox{The third column shows the SMT-LIB fragment resulting from the translation.}

\begin{small}
\setlength\tabcolsep{1pt}
\begin{longtable}[t!]{|p{13em}|p{13em}|p{15em}|}
\caption{OCL to SMT-LIB Translation Rules \label{tab:rules}}\\
\hline 
\centering \textbf{Node type possible within a processed NNF expression} & \multicolumn{1}{c|}{ \multirow{2}{*}{\textbf{Shape of the OCL expression}}}&   \multicolumn{1}{c|}{\multirow{2}{*}{\textbf{Corresponding  SMT-LIB formula}}}   \\ 
\hline 
\endfirsthead
\endhead

\ocl{PrimitiveLiteralExp}'s subtypes, \ocl{EnumLiteralExp}  & \variable{A constant literal \ocl{X}} & \ocl{X}  \\
\hline
\multirow{1}{*}{\ocl{IfExp}} & \ocl{if}(\variable{condition}) \ocl{then} \variable{body1}  \ocl{else} \variable{body2} \ocl{endif} &
\ocl{(ite} \textbf{toSMT-LIB}(\variable{condition}) \textbf{toSMT-LIB}(\variable{body1}) \textbf{toSMT-LIB}(\variable{body2})\ocl{) }
\\
\hline
\multirow{1}{*}{\ocl{VariableExp}} & \variable{A variable named \ocl{X}} & The name of the SMT-LIB variable that corresponds to  \ocl{X} \\
\hline
\multirow{1}{*}{\ocl{AttributeCallExp}}  & \variable{An access to an attribute named \ocl{X}} & The name of the SMT-LIB variable that corresponds to  \ocl{X} \\
\hline

 \multirow{1}{*}{\ocl{OperationCallExp}}  &  \variable{operand1} \variable{\ocl{op}} \variable{operand2}; Binary infix operations where  \variable{\ocl{op}} can be:  \ocl{and}, \ocl{or}, \ocl{<}, \ocl{>}, \ocl{<=}, \ocl{>=}, \ocl{+}, \ocl{-}, \ocl{*}, \ocl{/},
 \ocl{=}, or \ocl{<>} & \ocl{(}\variable{\ocl{op}}  \textbf{toSMT-LIB}(\variable{operand1})\newline  \textbf{toSMT-LIB}(\variable{operand2})\ocl{)} \\
\cline{2-3}

&  \ocl{not (}\variable{expression}\ocl{)} & \ocl{(not}  \textbf{toSMT-LIB}(\variable{expression})\ocl{)} \\
 \cline{2-3}

& \variable{\ocl{X}}\ocl{.}\variable{\ocl{op}}\ocl{()}; Calls to  OCL built-in operations that admit no parameters. Specifically,  \variable{\ocl{op}} can be:   \ocl{size}, \ocl{toInteger}, \ocl{toReal}, \ocl{round}, \ocl{ceil}, \ocl{abs}, or \ocl{floor}& 
 \ocl{(}\variable{\ocl{op}}  \textbf{toSMT-LIB}(\variable{X})\ocl{)}\newline
Except for the \ocl{abs} operation, which has a corresponding built-in function in SMT-LIB with the same name, the remaining operations need to be explicitly defined in SMT-LIB as illustrated below:

\begin{enumerate}
\item  \ocl{(define-fun sizeString ((a String)) Int (str.len a))}
\item  \ocl{(define-fun toInteger ((a String)) Int (str.to.int a))}
\item \ocl{(define-fun round ((a Real)) Real (ite (> (- a (floor a)) 0.5) (toReal (ceil a)) (toReal (floor a))))}
\item \ocl{(define-fun ceil ((x Real)) Int (+ (to\_int x) 1))}
\item \ocl{(define-fun floor ((x Real)) Int (to\_int x))}
\item \ocl{(define-fun toReal ((a Int)) Real (* a 1.0))}
\end{enumerate}

\\
\hline
& \variable{\ocl{X}}\ocl{.}\variable{\ocl{op}}\ocl{(\variable{\ocl{Y}})}; Calls to  OCL built-in operations that admit one parameter. Specifically,  \variable{\ocl{op}} can be:   \ocl{max}, \ocl{min}, \ocl{div}, \ocl{mod}, or \ocl{concat}& 
 \ocl{(}\variable{\ocl{op}}  \textbf{toSMT-LIB}(\variable{X})  \textbf{toSMT-LIB}(\variable{Y})\ocl{)}\newline
Except for \ocl{div} and \ocl{mod}, which have corresponding built-in functions in SMT-LIB (with the same names), the remaining operations need to be explicitly defined as illustrated below:
\begin{enumerate}
\item[7)]  \ocl{(define-fun max ((a Real)(b Real)) Real (ite (>= a b) a b))}
\item[8)] \ocl{(define-fun min ((a Real)(b Real)) Real (ite (<= a b) a b))}
\item [9)] \ocl{(define-fun concatString ((a String)(b String)) String (str.++ a b))}

\end{enumerate}

\\
 \cline{2-3}
&  \variable{\ocl{X}}\ocl{.}{\ocl{substring}}\ocl{(\variable{\ocl{Y}}, \variable{\ocl{Z}})} (this is the only OCL built-in operation on primitive variables that has two or more parameters)& 
 \ocl{(str.substr}  \textbf{toSMT-LIB}(\variable{X})  \textbf{toSMT-LIB}(\variable{Y}) \textbf{toSMT-LIB}(\variable{Z})\ocl{)}\\
  \cline{2-3}
 &  \variable{\ocl{X}}\ocl{.}\variable{user-defined operation O} \ocl{(}\variable{$\mathcal{P}$ parameters}\ocl{)} & 
 \ocl{(}\variable{O}  \variable{ \textbf{\textit{forEach}} ($\mathcal{P}_i \in \mathcal{P}$) \textbf{\textit{do}} \{\textbf{toSMT-LIB}({$\mathcal{P}_i$})\}}\ocl{)}\\
 \hline

\end{longtable}
\end{small}

\end{document}